\documentclass[%
 reprint,
 amsmath,amssymb,
 aps,
]{revtex4-2}

\usepackage{graphicx}
\usepackage{dcolumn}
\usepackage{bm}
\usepackage{xcolor}

\pagecolor{white}
\begin{document}

\preprint{APS/123-QED}

\title{Non-Perturbative Renormalization Group for Ising-Nematic Criticality: A Closed-Form Nonlocal Ansatz}

\author{Hyeon Jung Kim$^{1}$, Kyoung-Min Kim$^{1,2}$, and Ki-Seok Kim$^{1,2}$}

	\email[Hyeon Jung Kim: ]{hkim7218@postech.ac.kr}
	\email[Ki-Seok Kim: ]{tkfkd@postech.ac.kr}

\affiliation{$^{1}$Department of Physics, POSTECH, Pohang, Gyeongbuk 37673, Korea\\
$^{2}$Asia Pacific Center for Theoretical Physics, Pohang, Gyeongbuk 37673, Korea}

\date{\today}

\begin{abstract}
The two-dimensional metallic quantum critical problem is a long-standing puzzle that is widely believed to hold the key to resolving ubiquitous non-Fermi liquid behavior in strongly correlated electronic systems. In this study, we present a non-perturbative renormalization group (RG) analysis of the metallic Ising-nematic quantum critical point in two dimensions, formulated directly around an intrinsically nonlocal infrared (IR) boson propagator. Rather than treating the anomalous dynamical critical exponent $a$ as a fixed phenomenological parameter, we regard it as an intrinsic component of the fixed-point data to be determined from the internal consistency of the low-energy patch field theory under highly anisotropic scaling dimensions ($[k_0]=a+1$, $[k_x]=2$, $[k_y]=1$). While the leading two-loop diagrammatics vanish identically due to kinematic pole configurations, our three-loop evaluation reveals a profound structural asymmetry between the sectors: the fermion self-energy and Yukawa vertex receive non-vanishing logarithmic corrections, whereas the corresponding bosonic counter-term remains strictly zero. Consequently, we find that no self-consistent, intersecting fixed-point solution for the exponent $a$ exists within the three-loop truncation, failing to reproduce the physical value of $a \approx 1.85$ observed in quantum Monte Carlo simulations. We conjecture that the cross-linked topology of the four-loop boson self-energy diagrams is exactly marginal and yields the minimal, mandatory bosonic counter-term required to restore multi-sector self-consistency. Our framework establishes a rigid multi-loop matching scheme necessary to uniquely pin down the critical exponent, and uncovers a stable phase space for field anomalous dimensions.
\end{abstract}

\maketitle


\section{Introduction}

Landau's Fermi-liquid theory establishes the standard low-energy framework
for interacting fermions with a well-defined Fermi surface
\cite{Landau1957, Nozières1964}. Within this paradigm, elementary
excitations manifest as long-lived quasiparticles near the Fermi surface,
yielding a sharp and distinct peak in the electron spectral function
\cite{Mahan2000}. This canonical description breaks down, however, when
low-energy fermions become strongly coupled to collective fluctuations
\cite{Hertz1976, Millis1993}. Such strong coupling can destroy the
quasiparticle pole entirely, driving the system into a metallic state known
as a non-Fermi liquid \cite{Stewart2001, Lee2018, Loehneysen2007}.

Non-Fermi-liquid behavior emerges across a diverse range of quantum
many-body systems, typically driven by the coupling between gapless
fermionic excitations and a gapless bosonic mode
\cite{Holstein1973, Reizer1989}. This bosonic mode may arise either as an
order-parameter fluctuation near a quantum critical point or as a collective
mode that remains gapless in the infrared
\cite{Polchinski1994, Altshuler1994}. In systems with spontaneous symmetry
breaking, however, the mere existence of a Goldstone mode is not sufficient
to guarantee non-Fermi-liquid behavior \cite{Watanabe2014}. Its effect on the Fermi surface
depends on how the broken symmetry acts on low-energy quasiparticle states.
When the Goldstone mode is restricted to derivative couplings, its effective
interaction with low-energy fermions typically vanishes in the infrared. In
contrast, when the broken generator acts nontrivially on quasiparticle
momentum states, the associated Goldstone mode can retain a nonvanishing
low-energy coupling to the Fermi surface and can provide a route to
non-Fermi-liquid behavior. The core theoretical
challenge then lies in the mutual renormalization between the bosonic and
fermionic degrees of freedom: gapless fermion particle-hole excitations
induce nonlocal dynamics in the bosonic sector
\cite{Nayak1994, Halperin1993}, while the exchange of low-energy bosons
generates a nonanalytic fermion self-energy that overwhelms the bare
frequency term \cite{Chubukov2003, SSLee2009}. Consequently, the fermionic
and bosonic sectors become intrinsically entangled and cannot be treated
independently in the low-energy limit \cite{Sachdev2010prb}.

Analogous non-Fermi-liquid phenomena have been extensively explored in
spin-liquid systems \cite{Nayak1994b, Lee2005, Lee2017prx} and
in the context of nematic ordering within $d$-wave superconductors
\cite{Sachdev2008prb, Fritz2009}. These systems serve as prime examples of
low-energy fermions interacting strongly with gapless collective modes. In
this work, we focus on the metallic Ising-nematic quantum critical point,
where a small-momentum nematic order parameter couples directly to fermions
residing on the Fermi surface
\cite{Oganesyan2001, Sachdev2010prb, Kivelson2003}.

The Ising-nematic quantum critical point stands as one of the simplest
archetypes of quantum criticality intertwined with a Fermi surface. The
Ising-nematic order parameter characterizes the spontaneous breaking of the
discrete rotational symmetry of the Fermi surface while preserving its
translational symmetry \cite{Kivelson1998, Fradkin2010}. In the language of
Landau Fermi-liquid theory, this transition corresponds to a Pomeranchuk
instability in the angular momentum $l=2$ channel
\cite{Pomeranchuk1958, Quintanilla2006}. Because the order parameter carries
nearly zero momentum, it scatters fermions within the same local patch of
the Fermi surface \cite{Sachdev2010prb, Holder2012}. Thus, despite the
apparent simplicity of the Ising symmetry, the low-energy critical theory
remains deeply altered by the presence of gapless fermions \cite{Garst2010}.

A conventional approach to quantum criticality with a Fermi surface is the
Hertz-Millis construction, which integrates out the fermionic degrees of
freedom to yield an effective action solely for the bosonic order parameter
\cite{Hertz1976, Millis1993}. For the Ising-nematic quantum critical point
in two dimensions, however, this procedure proves fundamentally inadequate
\cite{Loehneysen2007,Lee2018,Abanov2003}. Because fermions remain gapless along the
entire Fermi surface, integrating them out generates singular, nonlocal
terms in the bosonic effective action
\cite{Sachdev2010prb, Chubukov2005}. A purely bosonic framework is therefore
incapable of providing a complete scaling description of the critical point;
a systematic low-energy analysis must explicitly retain both the
order-parameter fluctuations and the fermionic excitations
\cite{Senthil2004, Senthil2010prb}.

Many attempts have been made to construct a controlled theory of the
Ising-nematic quantum critical point. Large-$N$ generalizations offer one
potential organizing principle \cite{Polchinski1994, Altshuler1994}, yet
the large-$N$ limit becomes inherently singular when a Fermi surface is
coupled to a gapless boson \cite{SSLee2009}. In the two-patch approach, higher-loop corrections are organized around the one-loop dressed fermion and boson propagators, but the resulting expansion lacks a genuine small control parameter \cite{Sachdev2010prb}. In particular, the three-loop boson self-energy contains an $N^{3/2}$ contribution, which is parametrically larger than the bare $O(N)$ term. This enhancement arises because interpatch contributions evade the external-frequency bound on internal frequencies, allowing an $N$-dependent crossover region to dominate the integral and generate an additional factor of $N^{1/2}$ beyond genus counting. This anomaly causes the naive $1/N$ expansion to
break down, as an infinite class of diagrams contributes at the same order,
stripping the theory of systematic $1/N$ counting control
\cite{Chubukov2006, Mandal2015}. The Ising-nematic problem suffers from
this exact obstruction because higher-loop corrections are not suppressed
in the large-$N$ limit \cite{Fitzpatrick2013}.

To circumvent this, alternative controlled expansions have been proposed
\cite{Senthil2010prb, Dalidovich2013}. One prominent strategy involves
deforming the boson dynamics, thereby pairing the large-$N$ expansion
with an expansion in the boson dynamical exponent
\cite{Mahajan2013, Sur2014}. While this approach offers a controlled window
into certain non-Fermi-liquid systems, including nematic quantum critical
theories, alongside valuable constraints from patch-theory analyses and
Ward identities \cite{Metlitski2010b, Dupuis2021}, unambiguously
determining the physical infrared fixed point remains a formidable open
challenge \cite{Lee2017prx, Lunts2018}.

In this work, we chart a different course to overcome these long-standing
obstructions by introducing a novel nonperturbative framework based on a
closed-form nonlocal RG approach \cite{Dupuis2021, Kopietz2010}. Motivated
by recent quantum Monte Carlo simulations
\cite{Meng2017prx, Schattner2016, Dumitrescu2016}, we assume that the
infrared bosonic dynamics is intrinsically nonlocal from the outset. Rather
than starting from a conventional local Hertz-Millis bosonic action and
deriving nonlocality by integrating out the fermions, we introduce the
following ansatz for the inverse boson propagator:
\begin{equation}
\begin{gathered}
    D^{-1}(q) =
    \left(c_1 q_0^2 + c_2 q_y^2\right)^{a/2}
    + c_3 \frac{|q_0|}{|q_y|}, \\
    1 < a < 2.
\end{gathered}
\label{eq:ansatz}
\end{equation}
Here, the second term represents the conventional Landau-damping
contribution generated by the gapless Fermi surface. The first term,
characterized by a nontrivial exponent $a$, captures the emergent
nonlocal bosonic dynamics in the infrared limit \cite{Wang2024}.

The central objective of this study is to determine the exponent $a$
strictly from the internal fixed-point consistency of the low-energy theory.
In this framework, $a$ is not treated as a fixed phenomenological
parameter; instead, it is regarded as an intrinsic component of the
fixed-point data to be solved self-consistently. We thus reverse the
standard problem: starting from this nonlocal boson propagator, we
systematically evaluate the scaling structures of the fermion self-energy,
boson self-energy, and Yukawa vertex using a closed-form RG ansatz.
Crucially, this ansatz reveals a distinct structural asymmetry between the
fermionic and bosonic sectors upon multiloop truncation. We demonstrate that
while the fermion self-energy and Yukawa vertex receive nonvanishing
logarithmic corrections at the three-loop level, the corresponding bosonic
counterterm remains strictly zero. Consequently, no self-consistent,
intersecting fixed-point solution for the exponent $a$ exists within the
three-loop order. This fails to reproduce the phenomenological value of
$a \approx 1.85$ reported in recent sign-problem-free quantum Monte Carlo
(QMC) simulations~\cite{Schattner2016, Meng2017prx}.

This breakdown at lower-order truncations rigorously motivates the
necessity of higher-order diagrammatics~\cite{Lee2017prx, Lunts2018}.
Through a comprehensive scaling analysis, we establish that determining the
critical exponent $a$ inherently requires a combined multiloop matching
scheme. Specifically, one must couple the three-loop fermionic and vertex
sectors with the four-loop boson self-energy corrections to fully restore
the exact fixed-point structure.

The remainder of this paper is organized as follows. In Sec.~II, we derive
the effective IR action associated with our ansatz and extract the scaling
dimensions of the fields and couplings. In Sec.~III, we formulate the exact
RG equations and analyze the multiloop consistency conditions.

\section{IR Effective Action}

\begin{figure}[htbp]
  \centering
  \includegraphics[width=0.65\linewidth]{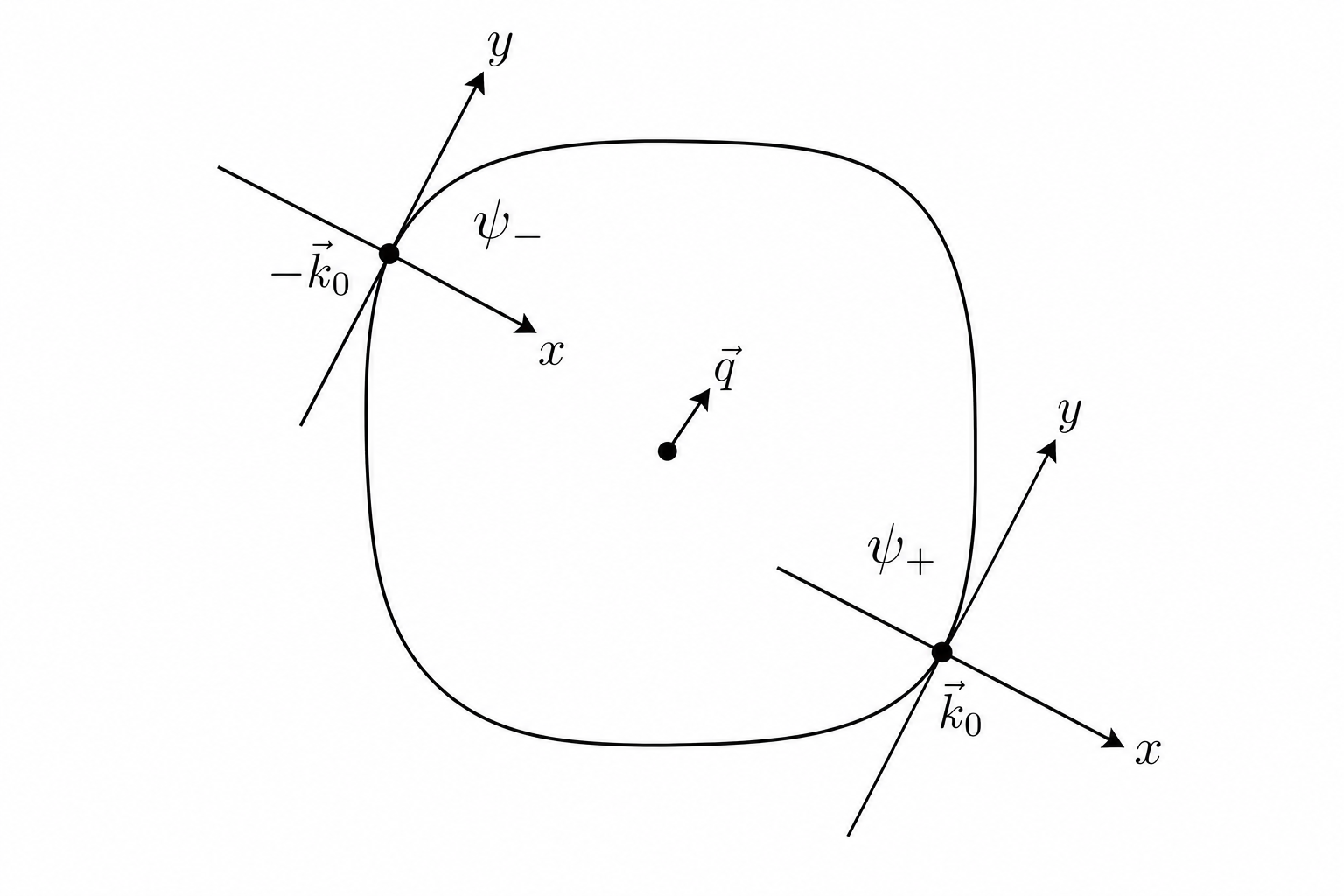}
  \caption{Schematic representation of the antipodal patches on the Fermi surface and the low-energy scattering processes governed by the effective action.}
  \label{fig:fermi_surface}
\end{figure}

\begin{figure}[htbp]
  \centering
  \begin{tabular}{cc}
    \includegraphics[width=0.5\linewidth]{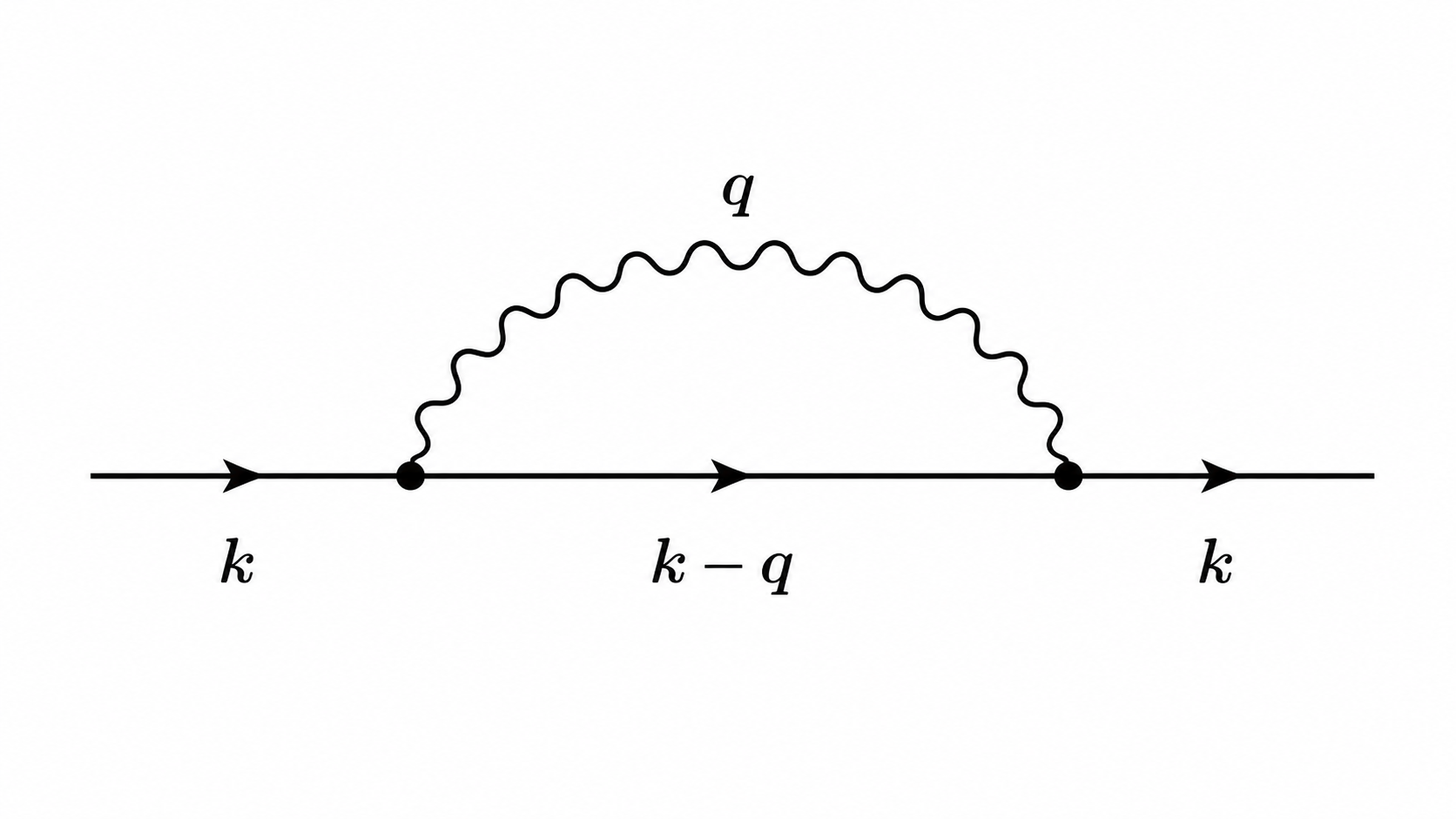} &
    \includegraphics[width=0.5\linewidth]{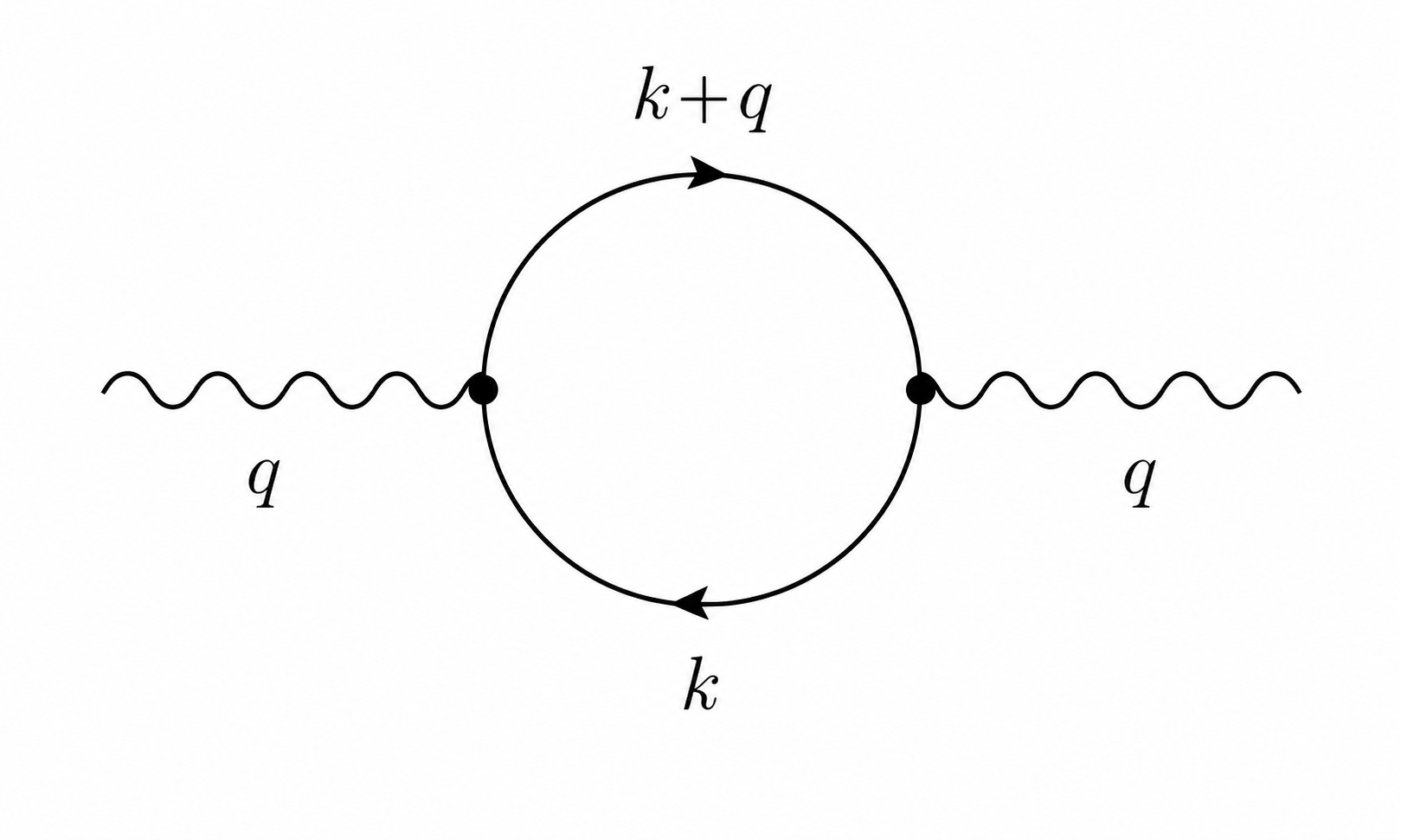} \\
    (a) Fermion self-energy& (b) Boson polarization
  \end{tabular}

  \vspace{10pt}
  
  \caption{The leading one-loop Feynman diagrams in the two-patch effective theory: (a) the one-loop fermion self-energy correction and (b) the one-loop boson polarization diagram, where solid lines represent dressed fermion propagators and wavy lines represent the nonlocal boson propagator.}
  \label{fig:one_loop_diagrams}
\end{figure}

We begin with the standard two-patch effective field theory for the
Ising-nematic quantum critical point. As shown in
FIG.~\ref{fig:fermi_surface}, the $x$-direction is taken to be normal to
the Fermi surface, whereas the $y$-direction is tangential to it.
Expanding the fermion dispersion near the two antipodal patches gives
\begin{equation}
    \epsilon_s(k)=s k_x+k_y^2,
\end{equation}
where $s=\pm$ denotes the patch index.
The corresponding low-energy Lagrangian density is given by
\begin{equation}
\begin{aligned}
\mathcal{L}
=&\sum_{s=\pm} \psi^{\dagger}_{s\sigma}
\left( \partial_{\tau} - is\partial_x - \partial_y^2 \right)
\psi_{s\sigma} \\
&+ g \sum_{s=\pm} \phi \psi^\dagger_s \psi_s
+ \frac{c}{2} (\partial_y \phi)^2
+ \frac{r}{2} \phi^2 .
\end{aligned}
\label{eq:lagrangian}
\end{equation}
where $\psi$ is a fermionic field, and the Ising-nematic order parameter is
represented by a bosonic field $\phi$, which couples to the local fermion
density on each patch via a Yukawa interaction.

At the quantum critical point, the boson couples strongly to gapless
particle-hole excitations on the Fermi surface, generating Landau damping
that profoundly alters the infrared dynamics. Motivated by the nonlocal
nature of the critical boson mode, we employ the dressed boson propagator
instead of its bare local counterpart. This coupling renders the corresponding
one-loop fermion self-energy, shown in FIG.~\ref{fig:one_loop_diagrams}(a),
singular in frequency. Incorporating this self-energy correction, the
RPA-dressed inverse fermion propagator takes the form
\begin{equation}
G^{-1}_{s}(k)
=
-ik_0 + s k_x + k_y^2
-
\frac{i g^2 \operatorname{sgn}(k_0)}
{2\pi c_3 \sin\left(\frac{2\pi}{a+1}\right)}
\left(
\frac{c_3 |k_0|}{c_2^{a/2}}
\right)^{\frac{2}{a+1}} .
\label{eq:RPA_dressed_fermion}
\end{equation}
For $1 < a < 2$, the exponent satisfies $2/(a+1) < 1$. Consequently, the
interaction-induced self-energy is more relevant in the infrared than the
bare frequency term, which constitutes the defining non-Fermi-liquid feature
of this low-energy theory. In the same regime, the one-loop polarization
shown in Fig.~\ref{fig:one_loop_diagrams}(b) is given by
\begin{equation}
\Pi^{1L}(q) = \frac{g^2}{4\pi}\frac{|q_0|}{|q_y|}.
\label{eq:RPA_dressed_boson}
\end{equation}
Matching the amplitude of this Landau-damping term to that in the boson
propagator ansatz, Eq.~\eqref{eq:ansatz}, gives
\begin{equation}
\frac{g^2}{4\pi}=c_3,
\qquad
g^2=4\pi c_3 .
\end{equation}

\begin{figure}[htbp]
  \centering
  \includegraphics[width=0.6\linewidth]{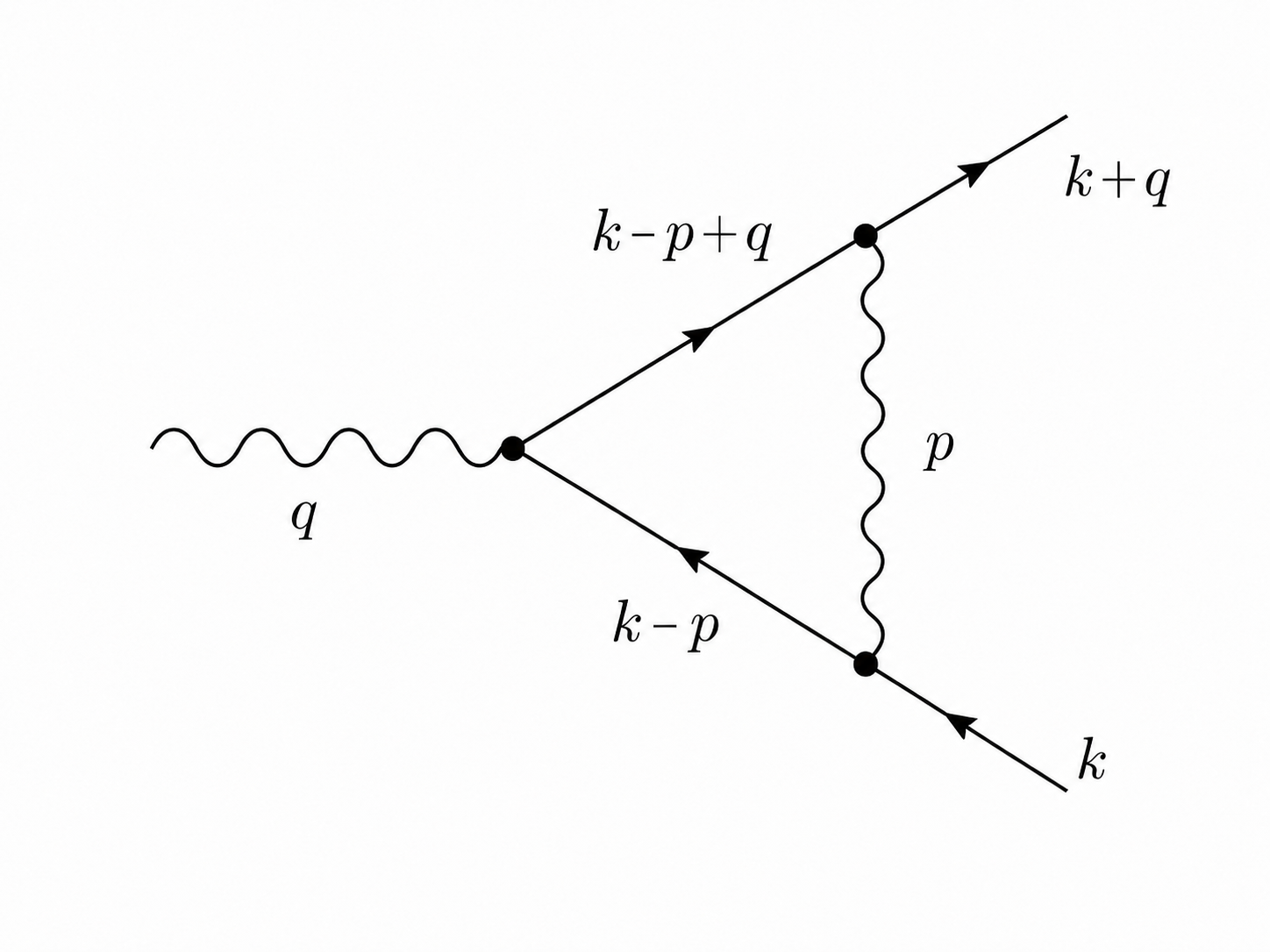}
  \caption{The one-loop Yukawa vertex correction diagram.}
  \label{fig:1-loop-vertex}
\end{figure}

FIG.~\ref{fig:1-loop-vertex} shows the one-loop correction to the Yukawa
vertex. In the infrared limit, this correction vanishes, since the fermionic
poles in the $k_x$ integration combine into a double pole whose residue gives
no additional infrared singularity. As a result, the one-loop vertex
correction does not introduce any additional infrared singularity, nor does it
alter the leading scaling behavior of the Yukawa interaction. Thus, at this
level of approximation, the Yukawa vertex may be consistently replaced by the
coupling constant $g$ in the low-energy effective action.

The scaling structure is determined by requiring the three dominant terms in the dressed fermion propagator, namely
\[
s k_x, \qquad k_y^2, \qquad \text{and} \qquad |k_0|^{2/(a+1)},
\]
to scale homogeneously. This requirement yields the anisotropic scaling dimensions:
\[
\boxed{
[k_0] = a + 1, \qquad [k_x] = 2, \qquad [k_y] = 1.
}
\]
With this choice, the inverse boson propagator also scales homogeneously, since $|q_y|^a$ and $|q_0|/|q_y|$ both possess a scaling dimension of $a$. The scaling dimensions of the fields are subsequently fixed by demanding the scale invariance of the quadratic fermion and boson actions, which yields:
\[
\boxed{
[\psi(k)] = -\frac{a+6}{2}, \qquad [\phi(k)] = -(a+2), \qquad [g] = 0.
}
\]
Crucially, the Yukawa coupling $g$ is marginal at tree level. The infrared theory is therefore not obtained by simply decoupling the fermions and bosons; rather, their mutual interaction remains an intrinsic part of the fixed-point theory.
Combining these ingredients, the corresponding low-energy effective action is written as
\begin{widetext}
\begin{equation}
\begin{aligned}
S_{\mathrm{eff}}
=&
\sum_{s=\pm} \int \frac{d^{3}k}{(2\pi)^3}\,
\psi^{\dagger}_{\sigma,s}(k)
\biggl[
s k_x + k_y^2
- \frac{i g^2 \operatorname{sgn}(k_0)}
{2\pi c_3 \sin\left(\frac{2\pi}{a+1}\right)}
\left(
\frac{c_3 |k_0|}{c_2^{a/2}}
\right)^{\frac{2}{a+1}}
\biggr]
\psi_{\sigma,s}(k)
\\
&+
\frac{1}{2}
\int \frac{d^{3}q}{(2\pi)^3}\,
\phi(-q)
\biggl[
c_2^{a/2}|q_y|^a
+
c_3\frac{|q_0|}{|q_y|}
\biggr]
\phi(q)
+
\sqrt{4\pi c_3}
\sum_{s=\pm}
\int \frac{d^{3}k}{(2\pi)^3}
\frac{d^{3}q}{(2\pi)^3}\,
\psi^\dagger_{\sigma,s}(k+q)\,
\phi(q)\,
\psi_{\sigma,s}(k).
\end{aligned}
\label{eq:effective_action}
\end{equation}
\end{widetext}
This effective action serves as the point of departure for the subsequent analysis. The exponent $a$ remains undetermined at this stage, and the central question is whether its value can be uniquely fixed by the self-consistency of the RG flow.


From this effective action, we first evaluate the exponent $a$ by imposing a self-consistency condition on the boson propagator. The objective is to compute the leading two-loop corrections to the boson self-energy using the dressed fermion propagators and to verify whether these corrections either reproduce the assumed nonlocal form of $D^{-1}(q)$ or generate an independent constraint that uniquely determines $a$.

\begin{figure}[htbp]
  \centering
  \begin{tabular}{cc}
    \includegraphics[width=0.5\linewidth]{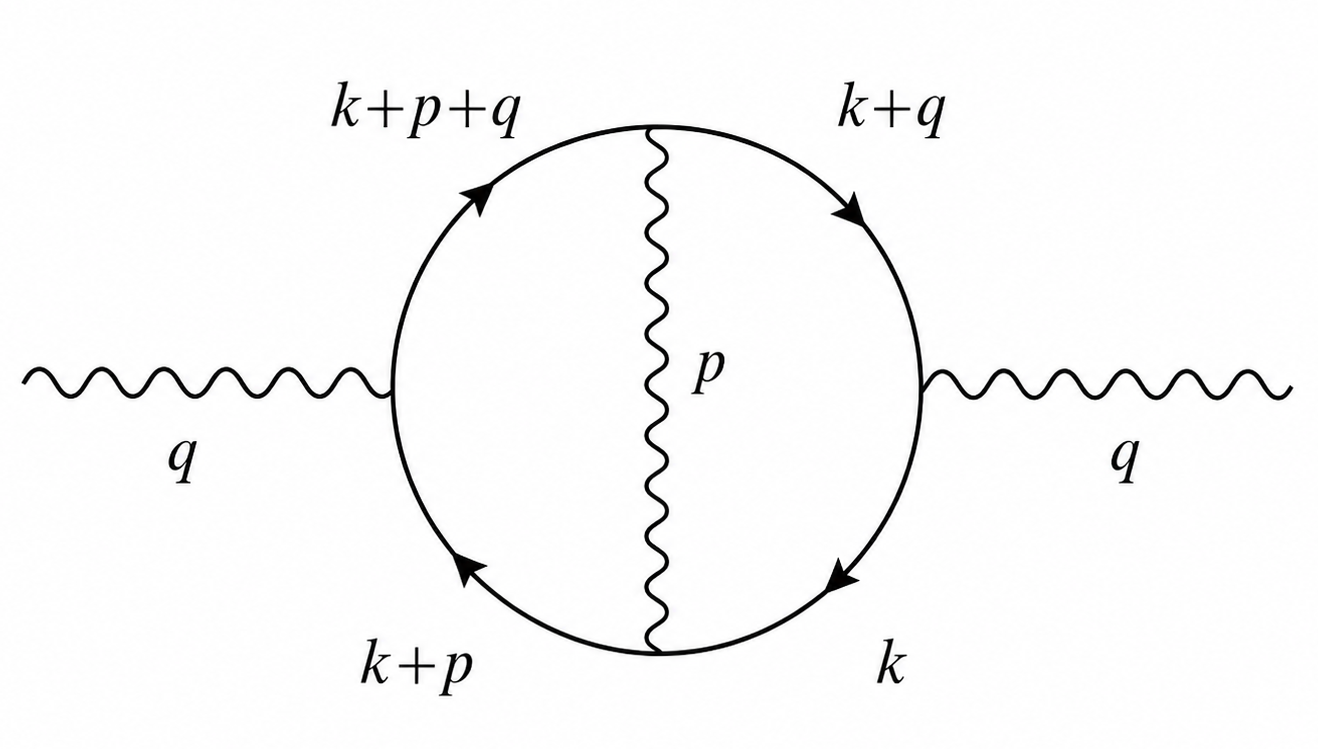} &
    \includegraphics[width=0.4\linewidth]{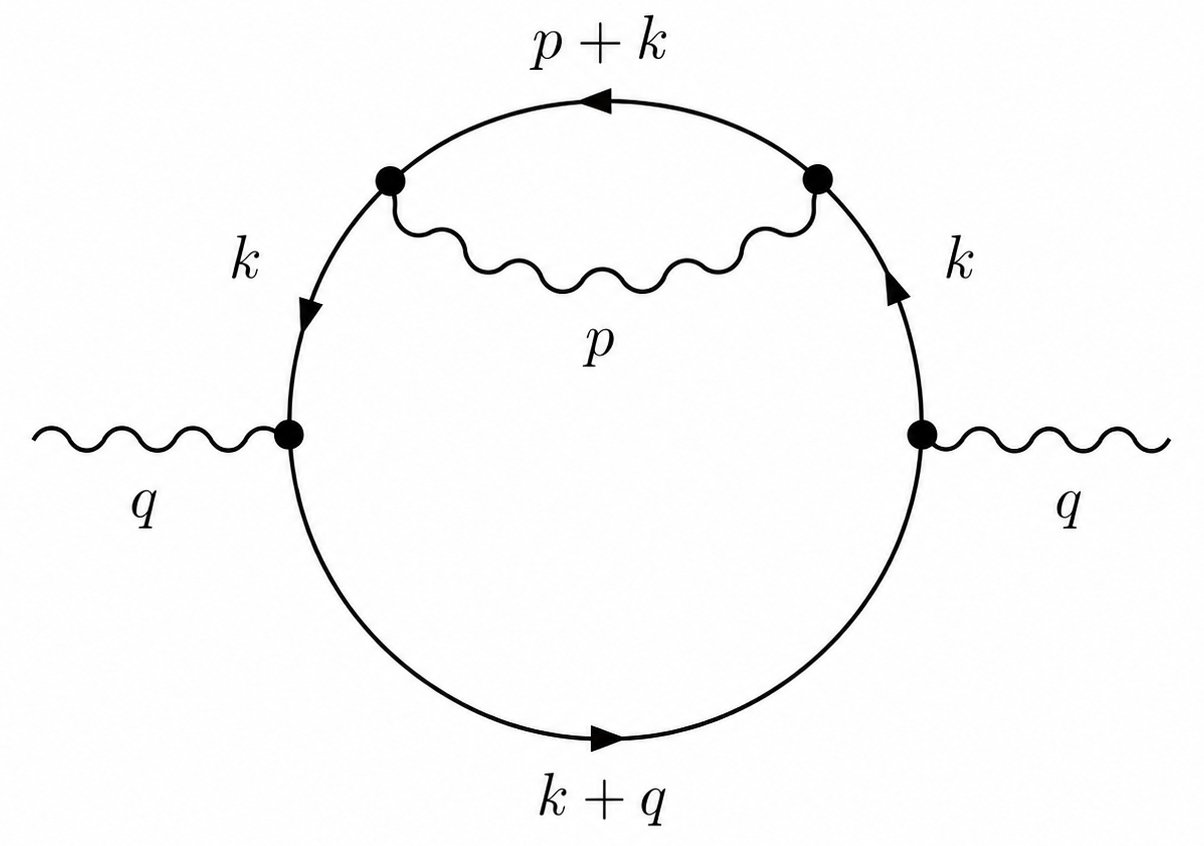} \\
    (a)  $\Pi^{2L}_1(q)$ & (b)  $\Pi^{2L}_2(q)$\\
  \end{tabular}
  \caption{The two-loop Feynman diagrams contributing to the boson self-energy:  (a) the Maki-Thompson vertex-type correction and (b) the diagram with a one-loop fermion self-energy insertion.}
  \label{fig:MT}
\end{figure}

The first two-loop contribution, corresponding to the Maki-Thompson diagram (FIG.~\ref{fig:MT} (a)), is given by
\begin{widetext}
\begin{equation}
\begin{aligned}
\Pi^{2L}_1(q)
=&\,
\frac{g^4}{(2\pi)^4}
\int dk_0\,dk_y\,dp_0\,dp_y\,
\frac{1}{c_2^{a/2}|p_y|^a+c_3 |p_0|/|p_y|}
\\
&\times
\biggl[
\frac{\Theta(k_0+q_0)-\Theta(k_0)}
{q_x+q_y^2+2k_yq_y
-i\left[\Sigma_1(k_0+q_0)-\Sigma_1(k_0)\right]}
\frac{\Theta(k_0+q_0+p_0)-\Theta(k_0+p_0)}
{q_x+q_y^2+2q_y(k_y+p_y)
-i\left[\Sigma_1(k_0+p_0+q_0)-\Sigma_1(k_0+p_0)\right]}
\biggr].
\end{aligned}
\label{eq:Pi_2L_1}
\end{equation}
\end{widetext}
where, $\Sigma_1(p)=2\text{sgn}(p_0)\Big(\frac{c_3}{c_2^{a/2}}\Big)^{2/(a+1)}|p_0|^{2/(a+1)}$. After performing the $k_x$ and $p_x$ integrations, the remaining integrand features two distinct denominators that depend linearly on $k_y$. The locations of these poles are determined by the self-energy differences:
\[
\Sigma_1(k_0+q_0)-\Sigma_1(k_0) \qquad \text{and} \qquad \Sigma_1(k_0+p_0+q_0)-\Sigma_1(k_0+p_0).
\]
The first difference scales as
\[
\begin{aligned}
&\Sigma_1(k_0+q_0)-\Sigma_1(k_0)\\
&=
C\biggl[
\operatorname{sgn}(k_0+q_0)
|k_0+q_0|^{2/(a+1)}-\operatorname{sgn}(k_0)
|k_0|^{2/(a+1)}
\biggr],
\end{aligned}
\]
Where, $C=2\Big(\frac{c_3}{c_2^{a/2}}\Big)^{2/(a+1)}$Within the kinematic domain constrained by the step functions, this difference shares the identical sign structure as $q_0$. A parallel argument applies to the second self-energy difference. Consequently, both $k_y$-poles are located within the same half of the complex plane. The integration contour can therefore be closed in the opposite half-plane, yielding a vanishing residue. This demonstrates that the first two-loop correction vanishes identically and does not constrain the value of $a$.

The second two-loop contribution can be simplified by identifying one internal boson line with the one-loop fermion self-energy insertion (FIG. \ref{fig:MT}(b)). This yields
\begin{widetext}
\begin{equation}
\Pi^{2L}_2(q)
=
-\frac{i g^2}{(2\pi)^2}
\int dk_0\,
\Sigma(k_0)
\left[
\Theta(k_0+q_0)-\Theta(k_0)
\right]
\int dk_y\,
\left[
\frac{1}{
q_x+q_y^2+2k_yq_y
-i\left(\Sigma_1(k_0+q_0)-\Sigma_1(k_0)\right)}
\right]^2 .
\label{eq:Pi_2L_2}
\end{equation}
\end{widetext}
In this case as well, all poles reside in the same half-plane, and the contour integration about $k_y$ yields zero.

\section{Renormalization Group Analysis}

The two-loop self-consistency analysis detailed above demonstrates that the exponent $a$ cannot be uniquely determined solely from the boson self-energy corrections at this order. The Maki-Thompson-type diagrams vanish identically because all internal poles reside in the same complex half-plane. Consequently, the critical exponent $a$ cannot be fixed by a simple lower-order self-consistency requirement. To address this issue, we formulate a systematic RG analysis in which $a$ is treated as an intrinsic fixed-point exponent within the nonlocal low-energy field theory.

The central objective is to construct an exact RG framework tailored for the
nonlocal IR theory. In this formulation, the exponent $a$ is kept arbitrary in
the initial RG ansatz, and the resulting flow equations are derived directly
for the nonlocal infrared structures alongside the Yukawa vertex. The physical
value of $a$ is subsequently determined only after enforcing the fixed-point
consistency conditions across the boson self-energy, the fermion self-energy,
and the vertex corrections. In this sense, the nonlocal propagator does not
serve as an external phenomenological input to a perturbative RG scheme; rather,
it constitutes an integral part of the fixed-point structure to be solved
self-consistently.

We establish our RG analysis starting from the effective action in
Eq.~\eqref{eq:effective_action}, where the Yukawa coupling is parameterized as
$g=\sqrt{4\pi c_3}$. At this stage, the value of $a$ is treated as an
undetermined parameter to be fixed by the RG consistency conditions.

To capture the non-Fermi-liquid dynamics, we define the anisotropic scaling transformations as:
\begin{equation}
    k_0 = k_0' e^{-(a+1)l}, \qquad k_x = k_x' e^{-2l}, \qquad k_y = k_y' e^{-l}.
\end{equation}
Under these transformations, the momentum and frequency cutoffs scale accordingly:
\begin{equation}
{\footnotesize
\Lambda_y(l)=\Lambda_{y,0}e^{-l},\quad
\Lambda_x(l)=\Lambda_{x,0}e^{-2l},\quad
\Lambda_0(l)=\Lambda_{0,0}e^{-(a+1)l}
}
\label{eq:cutoff_scaling}
\end{equation}
Equivalently, these cutoff scaling Relations imply:
\begin{equation}
    \Lambda_x \sim \Lambda_y^2, \qquad \Lambda_0 \sim \Lambda_y^{a+1}, \qquad \text{and} \qquad l = -\ln\left(\frac{\Lambda_y}{\Lambda_{y,0}}\right).
\end{equation}
The corresponding field rescalings are chosen to be:
\begin{equation}
    \psi(k) = \psi(k') \exp\left[ \frac{1}{2} \int_0^l dl'\, (a + 6 - \eta_f) \right],
\end{equation}
and
\begin{equation}
    \phi(k) = \phi(k') \exp\left[ \frac{1}{2} \int_0^l dl'\, (2a + 4 - \eta_b) \right],
\end{equation}
where $\eta_f$ and $\eta_b$ denote the anomalous dimensions of the fermion and boson fields, respectively.

We parameterize the loop corrections to the inverse fermion propagator, the inverse boson propagator, and the Yukawa vertex as follows:
\begin{equation}
    G^{-1}(k) = G^{-1}_0(k) - \Sigma(k),
\end{equation}
\begin{equation}
    D^{-1}(q) = D^{-1}_0(q) - \Pi(q),
\end{equation}
and
\begin{equation}
    g' = \Gamma g,
\end{equation}
where $G_0$ and $D_0$ represent the underlying RPA-dressed propagators Eq.\ref{eq:RPA_dressed_fermion}, and Eq.\ref{eq:RPA_dressed_boson}.

We now introduce our central hypothesis: the closed-form ansatz for the differential RG corrections, expressed as
\begin{equation}
    \frac{d\Pi(q)}{dl} = A(a) \biggl[ c_2^{a/2}|q_y|^{a} + c_3\frac{|q_0|}{|q_y|} \biggr],
    \label{eq:ansatz_boson}
\end{equation}
\begin{equation}
    \frac{d\Sigma(k)}{dl} = B(a) \biggl[ -\frac{2i\operatorname{sgn}(k_0)}{\sin\left(\frac{2\pi}{a+1}\right)} \left(\frac{c_3}{c_2^{a/2}}\right)^{\frac{2}{a+1}} |k_0|^{\frac{2}{a+1}} + \left(k_x + s k_y^2\right) \biggr],
    \label{eq:ansatz_fermion}
\end{equation}
and
\begin{equation}
    \frac{d\Gamma}{dl} = C(a) \sqrt{4\pi c_3}.
    \label{eq:ansatz_vertex}
\end{equation}

This closed-form formulation constitutes the core building block of our RG analysis. It asserts that the interaction-driven loop corrections admit an exact parameterization utilizing the identical nonlocal infrared structures that define the effective action itself. Consequently, the high-dimensional functional RG flow is reduced to a set of ordinary differential equations for the coefficient functions $A(a)$, $B(a)$, and $C(a)$. The mathematical validity and consistency of this closed-form ansatz are verified in the Appendix A, where we explicitly evaluate the scaling behavior of the higher-loop diagrams.

After integrating out a thin momentum shell and rescaling back to the original cutoff, the effective action at the scale $l+dl$ is transformed into:
\begin{widetext}
\begin{equation}
\begin{aligned}
S_{l+dl}
=&
\sum_{s=\pm}
\int \frac{d^{3}k'}{(2\pi)^3}\,
\psi'^{\dagger}_{s,\sigma}(k')\,
(1-\eta_f dl)(1-Bdl)
\biggl[
s k_x' + k_y'^2
-
\frac{2i\operatorname{sgn}(k_0')}
{\sin\left(\frac{2\pi}{a+1}\right)}
\left(
\frac{c_3 |k_0'|}{c_2^{a/2}}
\right)^{\frac{2}{a+1}}
\biggr]
\psi'_{s,\sigma}(k')
\\[0.5em]
&+
\frac{1}{2}
\int \frac{d^{3}q'}{(2\pi)^3}\,
\phi(-q')\,
(1-\eta_b dl)(1-Adl)
\biggl[
c_2^{a/2}|q_y'|^a
+
c_3\frac{|q_0'|}{|q_y'|}
\biggr]
\phi(q')
\\[0.5em]
&+
g(1+Cdl)(1-\eta_f dl)
\left(1-\frac{\eta_b}{2}dl\right)
\sum_{s=\pm}
\int \frac{d^{3}k'}{(2\pi)^3}
\frac{d^{3}q'}{(2\pi)^3}\,
\psi'^{\dagger}_{s,\sigma}(k'+q')\,
\phi(q')\,
\psi'_{s,\sigma}(k') .
\end{aligned}
\label{eq:S_l_plus_dl}
\end{equation}
\end{widetext}
where the coupling coefficients $c_2$ and $c_3$ are allowed to flow implicitly with the RG scale $l$.

By demanding that the renormalized action preserves its original functional form under the RG flow, we obtain the matching conditions:
\begin{equation}
B(a) = -\eta_f, \qquad C(a) = -\frac{1}{2}A(a) + \eta_f.
\end{equation}
Simultaneously, the flow equations for the bosonic coefficients $c_2$ and $c_3$ are derived as:
\begin{equation}
\frac{dc_2}{dl} = -\frac{2}{a}(\eta_b + A)c_2, \qquad \frac{dc_3}{dl} = -(\eta_b + A)c_3.
\end{equation}
At a physically stable quantum fixed point, these coefficients must become scale-invariant ($dc_2/dl = dc_3/dl = 0$). This stationarity condition enforces:
\begin{equation}
A(a) = -\eta_b, \qquad B(a) = -\eta_f, \qquad C(a) = \frac{1}{2}\eta_b + \eta_f.
\end{equation}
These relations establish the fundamental fixed-point consistency conditions
governing the nonlocal low-energy effective theory. At this stage, the number
of independent equations matches the number of unknown critical exponents,
$(\eta_f,\eta_b,a)$, so that the fixed-point conditions can be closed in
principle. This matching is not accidental; it follows from the amplitude
matching of the Yukawa coupling $g$, the scaling relation between the bosonic
and fermionic sectors induced by the nonlocal exponent $a$, and the ansatz in
Eqs.~(\ref{eq:ansatz_boson})--(\ref{eq:ansatz_vertex}).

As demonstrated by the explicit two-loop calculation, no logarithmic corrections emerge to renormalize the critical exponent $a$ at this order. Within the two-loop approximation, the system yields:
\begin{equation}
A(a) = B(a) = C(a) = 0
\end{equation}
for an arbitrary value of $a$, leaving the fixed-point exponent entirely unconstrained.

At the three-loop level, both the fermion self-energy and the Yukawa vertex corrections receive nontrivial logarithmic contributions. We parameterize these higher-order corrections as:
\begin{equation}
B(a) = (a+1)\mathcal{J}_{\mathrm{tot}}(a), \qquad C(a) = (a+1)\mathcal{F}_{\mathrm{tot}}(a).
\end{equation}

\begin{figure}[htbp]
\centering
\includegraphics[width=1.0\linewidth]{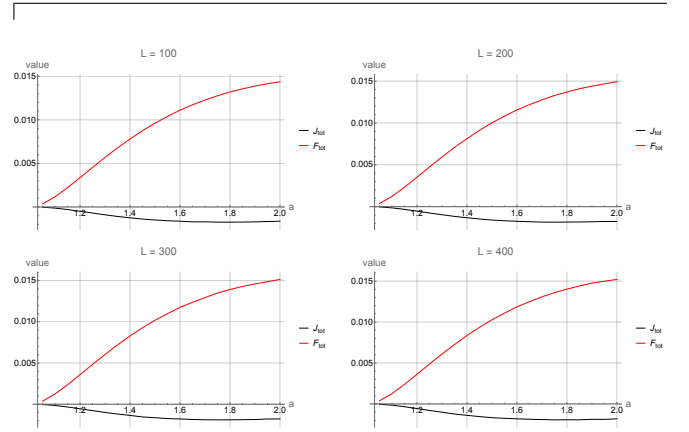}
\caption{Numerical evaluation of the $\mathcal{J}_{tot}(a)$ and $\mathcal{F}_{tot}(a)$ , calculated at the dimensionless infrared scale $L = (c_3/c_2^{a/2})\Lambda_0$. The plot visually demonstrates the complete absence of an intersecting self-consistent fixed-point solution within the three-loop order, signaling the breakdown of the three-loop truncation and the necessity of four-loop bosonic corrections.}
\label{fig:vertex_fermion_correction}
\end{figure}

Crucially, the boson self-energy sector does not yet generate the corresponding logarithmic divergence required to renormalize the nonlocal boson propagator at three-loop order. Thus, within the scope of the three-loop analysis, we find:
\begin{equation}
A(a) = 0.
\end{equation}

This imbalance leads to a striking and critical consequence for the theory. As illustrated in the numerical plot of $B(a)$ , and $C(a)$ (FIG.~\ref{fig:vertex_fermion_correction}), no self-consistent solution for the exponent $a$ exists within the three-loop truncation. Despite explicitly incorporating the intricate three-loop fermion self-energy corrections $B(a)$, the coupled fixed-point equations yield no mathematical intersection for any physical value of $a$ in the regime $1 < a < 2$. Consequently, the phenomenological value of $a \approx 1.85$ observed in recent quantum Monte Carlo simulations \cite{Meng2017prx} cannot be reproduced or sustained at this order.

\begin{figure}[htbp]
  \centering
  \includegraphics[width=0.6\linewidth]{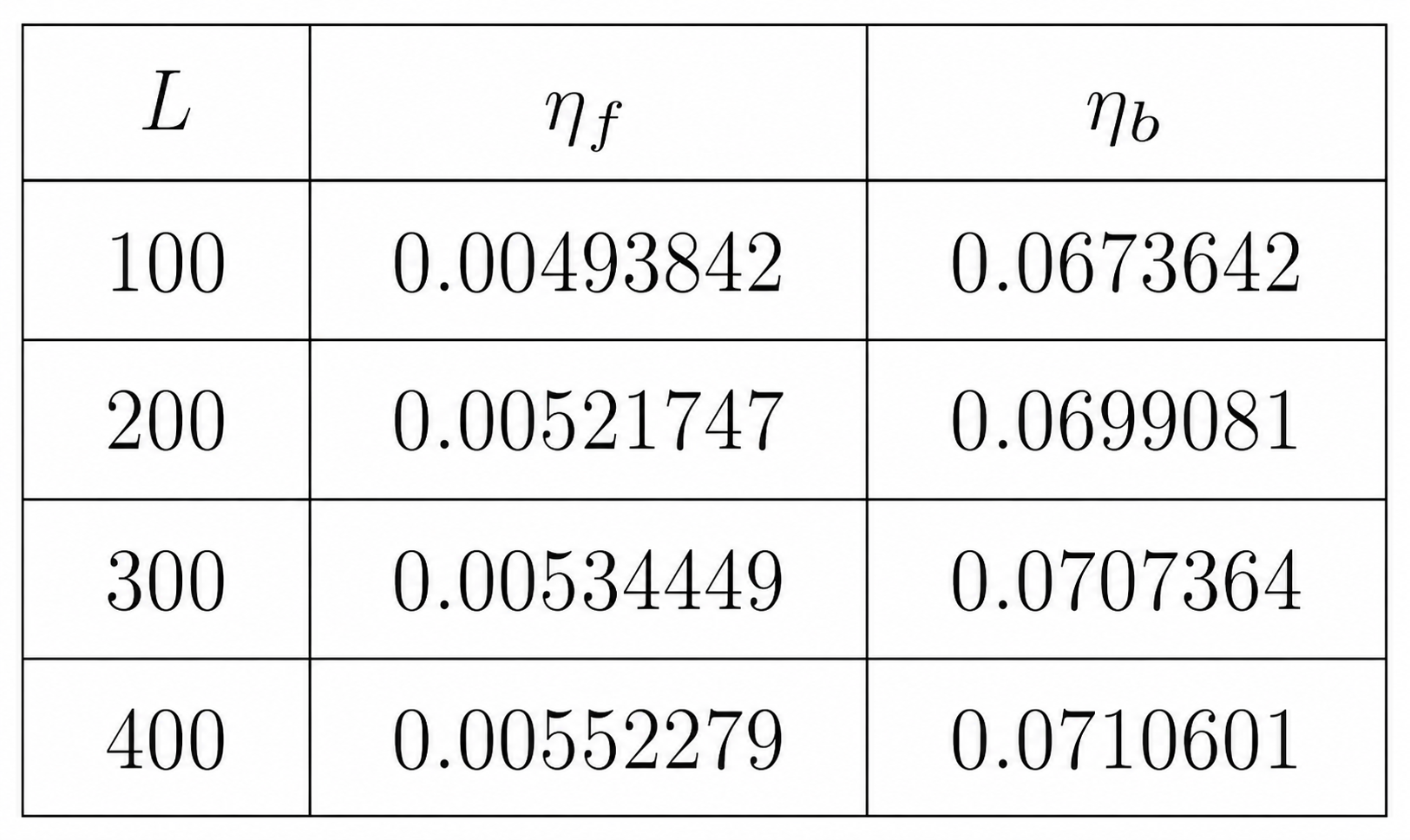}
  \caption{The values of $\eta_f$ and $\eta_b$ required for the RG equations to be
satisfied at the phenomenological value $a=1.85$. Here, $A$ is not obtained
from an explicit four-loop boson self-energy calculation, but is instead
treated as the constraint required to close the RG equations at this value
of $a$.}
  \label{fig:anomalous_dimensions}
\end{figure}

This complete absence of a three-loop fixed point provides a compelling, rigorous justification for advancing to higher-order diagrammatics. Because the three-loop bosonic counter-term vanishes ($A(a)=0$), the RG equations are structurally incapable of balancing the fermionic and vertex corrections. Therefore, we conjecture that the ultimate resolution of the critical exponent $a$ resides in the four-loop boson self-energy corrections. The boson anomalous dimension $\eta_b$, driven by the leading four-loop logarithmic divergences, is expected to generate a non-vanishing $A(a) \neq 0$\cite{Holder2015}.  As shown in FIG.~\ref{fig:anomalous_dimensions}, one may use the fixed-point condition to infer the value of the four-loop bosonic self-energy coefficient $A(a)$. However, the inferred values are not consistent with the infrared causality condition at $a=1.85$. In particular,
\begin{equation}
C(a) \equiv \frac{1}{2}\eta_b+\eta_f > 0
\end{equation}
implies that the Yukawa vertex diverges in the IR limit.

Crucially, within this strongly coupled infrared limit, the Yukawa interaction remains unsuppressed. The structural stability of the full loop corrections restricts the permissible phase space of the anomalous dimensions to:
\begin{equation}
    -a < \eta_b < 4, \qquad -2 < \eta_f < -\frac{\eta_b}{2},
\end{equation}
for $1 < a < 2$. This bound comfortably allows the fermion and boson anomalous dimensions to possess opposite signs, a non-trivial scaling phenomenon highly reminiscent of quantum electrodynamics (QED).

In this framework, the physical value ($a \approx 1.85$) \cite{Meng2017prx} should not be regarded as being determined by a truncation that combines the three-loop fermion self-energy, the three-loop Yukawa vertex correction, and the leading four-loop boson self-energy. Although this truncated system formally closes the fixed-point equations, the resulting solution is unphysical, indicating that still higher-order loop corrections are required for a consistent determination of $a$.

\section{Summary and Discussion}
\label{sec:discussion}

In this work, we have investigated the non-Fermi-liquid behavior at the metallic Ising-nematic quantum critical point within a non-perturbative RG framework. Departing from the conventional Hertz-Millis construction, we formulated the RG scheme directly around an intrinsically nonlocal infrared boson propagator. By treating the critical exponent $a$ as an internal fixed-point datum rather than as a fixed phenomenological input, we examined the structural self-consistency of the strongly coupled fermionic and bosonic sectors under the anisotropic scaling dimensions $[k_0]=a+1$, $[k_x]=2$, and $[k_y]=1$.

The central result of our analysis is the structural insufficiency of the low-order perturbative expansion. The leading two-loop diagrammatics vanish identically because of the kinematic arrangement of poles in the complex momentum plane. At three-loop order, however, the fermionic and bosonic sectors behave asymmetrically. The fermion self-energy and Yukawa vertex acquire non-vanishing logarithmic corrections, $\mathcal{J}_{\mathrm{tot}}(a)\neq 0$ and $\mathcal{F}_{\mathrm{tot}}(a)\neq 0$, whereas the corresponding bosonic counter-term remains zero, $A(a)=0$. Consequently, the three-loop truncation does not admit a self-consistent fixed-point solution for the exponent $a$. This demonstrates that the three-loop analysis alone is insufficient to determine the infrared fixed point and, in particular, cannot account for the physical value $a\simeq 1.85$ observed in quantum Monte Carlo simulations \cite{Meng2017prx}.

This failure of the three-loop truncation motivates the inclusion of
higher-order multiloop corrections. In particular, the leading
nonvanishing bosonic contribution is expected to arise from four-loop boson
self-energy diagrams. Once this contribution is included, the RG equations
can be formally closed by treating the four-loop bosonic coefficient
$A(a)$ as the missing constraint. However, at the phenomenological value
$a\simeq 1.85$, the anomalous dimensions inferred from this minimal
extension violate the infrared consistency condition. Specifically,
$C(a)\equiv \eta_b/2+\eta_f>0$, implying an infrared-divergent Yukawa
vertex. Thus, the leading four-loop bosonic restoration produces a formally
closed but physically unacceptable fixed point. A consistent determination
of $a$ therefore requires higher-order loop corrections beyond this minimal
four-loop extension.

Nevertheless, the low-energy effective theory still imposes a nontrivial
and constrained window for the anomalous dimensions. Requiring the bosonic
sector to remain infrared well-defined and the Yukawa vertex to avoid an
infrared divergence gives $-a<\eta_b<4$ and
$-2<\eta_f<-\eta_b/2$. These inequalities show that, even though the
minimal four-loop closure is not sufficient to determine a physical fixed
point, the fixed-point structure is not arbitrary. Rather, any consistent
solution must lie within a restricted region of anomalous dimensions.

Looking forward, the RG strategy developed here suggests a possible
direction for extending nonlocal RG methods to a broader class of quantum
field theories. The main outcome of the present work is not merely the
identification of a missing higher-loop correction in the Ising-nematic
problem, but the formulation of a nonlocal RG scheme in which the infrared
boson propagator and the critical exponent are treated self-consistently
within the fixed-point structure. Applying similar ideas to other systems
with intrinsically nonlocal propagators may clarify whether analogous
constraints on anomalous dimensions and loop-order hierarchies arise more
generally. In this sense, the present analysis should be viewed as
proposing one possible route toward nonlocal RG treatments of strongly
coupled quantum critical systems, rather than as providing a complete
framework for them.

An experimental system relevant to metallic nematic criticality is FeSe$_{1-x}$S$_x$. Elastoresistance measurements have provided evidence for a nematic quantum critical point without accompanying antiferromagnetic order~\cite{Hosoi2016}, and high-field transport measurements have revealed temperature-linear resistivity near this point~\cite{Licciardello2019}. These observations provide experimental context for the problem studied here, although they do not establish the specific nonlocal kernel or anomalous exponent considered in this work. Coupling between electronic nematicity and lattice strain can also modify the asymptotic low-energy behavior~\cite{Paul2017}.

\begin{acknowledgments}
K.-S. K. was supported by the Ministry of Education, Science, and Technology (Grant No. RS-2024-00337134) of the National Research Foundation of Korea (NRF). K.-M.K. was supported by an appointment to the JRG Program at the APCTP through the Science and Technology Promotion Fund and Lottery Fund of the Korean Government. We appreciate insightful discussions with J.-M. Bok, S.-J. Yoo, L. J. F. Sese, and S. Park.
\end{acknowledgments}

\appendix

\begin{widetext}
\section{General Scaling Analysis of Dressed Loop Diagrams}
\label{app:general_scaling}

In this Appendix, we present a general proof showing that the multi-loop diagrammatic corrections within our nonlocal RG framework are intrinsically independent of the individual bosonic tracking coefficients $c_2$ and $c_3$. We explicitly evaluate the scaling factor for an arbitrary loop order and demonstrate that the high-order perturbation series lacks a conventional, artificial small control parameter.


\subsection{Dressed Propagators and General Amplitudes}
\begin{figure}[htbp]
  \centering
  \includegraphics[width=0.5\linewidth]{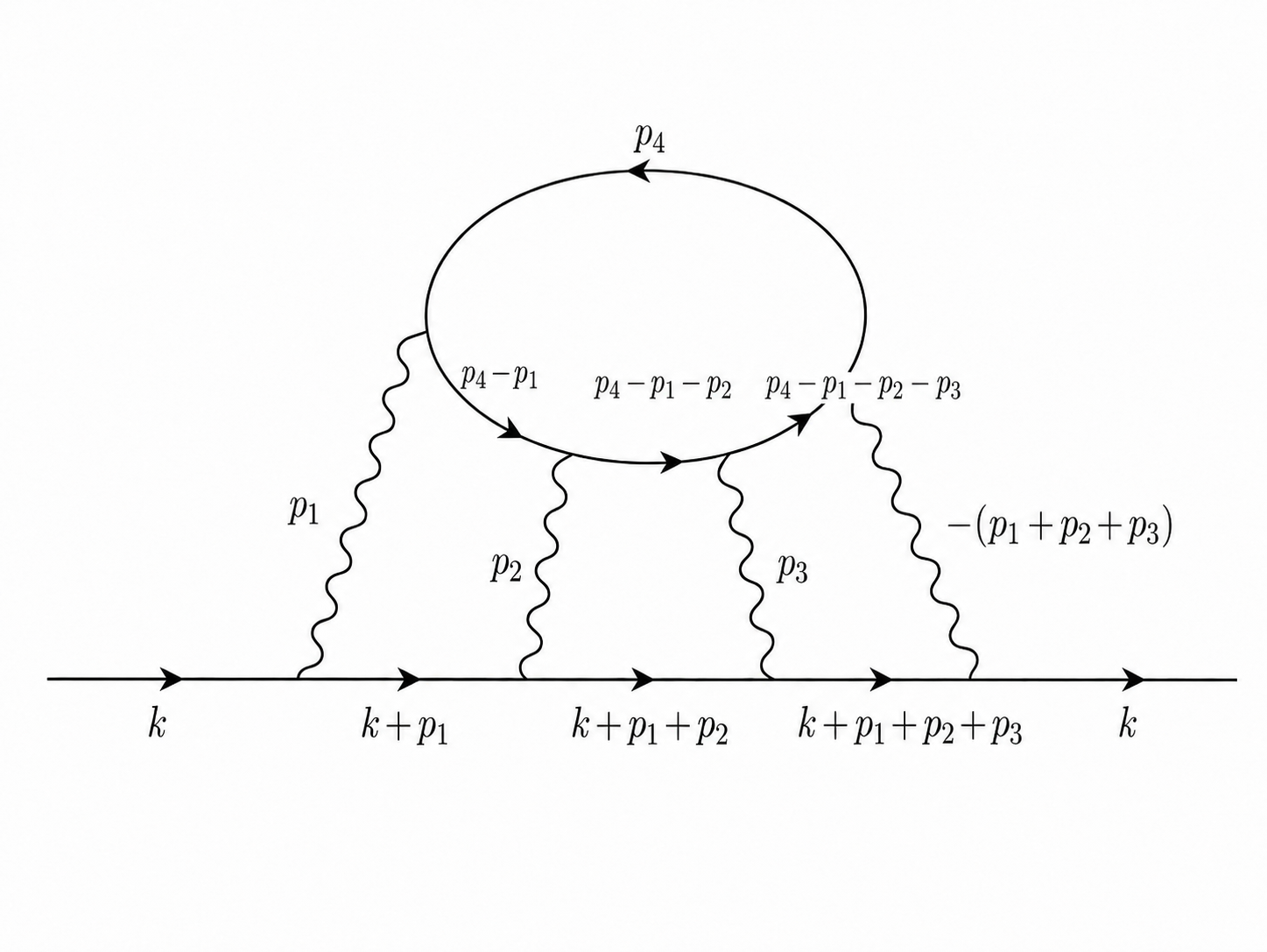}
  \caption{A example diagram}
  \label{fig:example_loop}
\end{figure}
We begin by summarizing the fundamental building blocks of the diagrammatic expansion. The dressed non-local boson propagator is defined as:
\begin{equation}
    D(q)^{-1} = \left(c_1 q_0^2 + c_2 q_y^2\right)^{a/2} + c_3 \frac{|q_0|}{|q_y|}, \qquad 1 < a < 2,
\end{equation}
and the corresponding RPA-dressed inverse fermion propagator takes the form:
\begin{equation}
    G^{-1}(k) = k_x + k_y^2 - \frac{2i\operatorname{sgn}(k_0)}{\sin\left(\frac{2\pi}{a+1}\right)} \left( \frac{c_3 |k_0|}{c_2^{a/2}} \right)^{\frac{2}{a+1}} \equiv k_x + k_y^2 - i\Sigma_1(k_0).
\end{equation}

To illustrate the scaling property, consider a specific multi-loop skeleton diagram in FIG.\ref{fig:example_loop} evaluated at zero external fermion momentum ($k=0$), whose amplitude $I$ can be formally written as:
\begin{equation}
\begin{split}
    I \sim c_3^{4} \int \prod_{r=1}^4 d^3p_r & \left( \prod_{j=1}^3 \frac{1}{c_2^{a/2}|p_{j,y}|^a + c_3 \frac{|p_{j,0}|}{|p_{j,y}|}} \right) \frac{1}{c_2^{a/2} \left| \sum_{n=1}^3 p_{n,y} \right|^a + c_3 \frac{\left| \sum_{n=1}^3 p_{n,0} \right|}{\left| \sum_{n=1}^3 p_{n,y} \right|}} \\
    & \times \frac{1}{p_{4,x} + p_{4,y}^2 - i\Sigma_1(p_{4,0})} \frac{1}{p_{1,x} + p_{1,y}^2 - i\Sigma_1(p_{1,0})} \\
    & \times \frac{1}{(p_{1,x} + p_{2,x}) + (p_{1,y} + p_{2,y})^2 - i\Sigma_1(p_{1,0} + p_{2,0})} \\
    & \times \frac{1}{(p_{1,x} + p_{2,x} + p_{3,x}) + (p_{1,y} + p_{2,y} + p_{3,y})^2 - i\Sigma_1(p_{1,0} + p_{2,0} + p_{3,0})} \\
    & \times \frac{1}{(p_{4,x} - p_{1,x}) + (p_{4,y} - p_{1,y})^2 - i\Sigma_1(p_{4,0} - p_{1,0})} \\
    & \times \frac{1}{(p_{4,x} - p_{1,x} - p_{2,x}) + (p_{4,y} - p_{1,y} - p_{2,y})^2 - i\Sigma_1(p_{4,0} - p_{1,0} + p_{2,0})} \\
    & \times \frac{1}{(p_{4,x} - \sum_{n=1}^3 p_{n,x}) + (p_{4,y} - \sum_{n=1}^3 p_{n,y})^2 - i\Sigma_1(p_{4,0} - p_{1,0} - p_{2,0} - p_{3,0})}.
\end{split}
\label{eq:example}
\end{equation}


\subsection{Anisotropic Coordinate Transformations and Prefactor Scaling}

To systematically isolate the $c_2$ and $c_3$ dependence from the integrands, we implement a homogeneous anisotropic coordinate transformation for all loop momenta ($j=1,2,3,4$):
\begin{equation}
    p_{j,0} \rightarrow p_{j,0}, \qquad p_{j,x} \rightarrow \left(\frac{c_3}{c_2^{a/2}}\right)^{\frac{2}{a+1}} p_{j,x}, \qquad p_{j,y} \rightarrow \left(\frac{c_3}{c_2^{a/2}}\right)^{\frac{1}{a+1}} p_{j,y}.
\end{equation}
Under this coordinate rescaling, the Jacobian for each loop momentum integration volume becomes:
\begin{equation}
    J_j = \left(\frac{c_3}{c_2^{a/2}}\right)^{\frac{3}{a+1}}.
\end{equation}
Concurrently, the inverse boson and fermion propagators map onto dressed dimensionless structures according to:
\begin{equation}
    \frac{1}{c_2^{a/2}|p_{j,y}|^a + c_3 \frac{|p_{j,0}|}{|p_{j,y}|}} \longrightarrow \frac{1}{c_3} \left(\frac{c_3}{c_2^{a/2}}\right)^{\frac{1}{a+1}} \frac{1}{|p_{j,y}|^a + \frac{|p_{j,0}|}{|p_{j,y}|}},
\end{equation}
\begin{equation}
    \frac{1}{p_{j,x} + p_{j,y}^2 - i\Sigma(p_{j,0})} \longrightarrow \left(\frac{c_3}{c_2^{a/2}}\right)^{-\frac{2}{a+1}} \frac{1}{p_{j,x} + p_{j,y}^2 - \frac{2i\operatorname{sgn}(p_{j,0})}{\sin\left(\frac{2\pi}{a+1}\right)} |p_{j,0}|^{\frac{2}{a+1}}}.
\end{equation}

We now generalize the prefactor counting carried out explicitly for the
example diagram in Eq.~\eqref{eq:example} to an arbitrary Feynman diagram containing
$L$ independent loop momenta, $V$ Yukawa vertices, $I_b$ internal boson
lines, and $I_f$ internal fermion lines.
\begin{equation}
\begin{split}
     I & \sim c_3^{V/2} \int \prod_{r=1}^L d^3p_r \left( \prod_{l=1}^{I_b} \frac{1}{c_2^{a/2}|p_{l,y}|^{a} + c_3 \frac{|p_{0,l}|}{|p_{l,y}|}} \right) \left( \prod_{m=1}^{I_f} \frac{1}{q_{m,x} + q_{m,y}^2 - i\Sigma_1(q_{m,0})} \right) \\
     & \equiv F \int \prod_{r=1}^L d^3p_r \left( \prod_{l=1}^{I_b} \frac{1}{|p_{l,y}|^{a} + \frac{|p_{0,l}|}{|p_{l,y}|}} \right) \left( \prod_{m=1}^{I_f} \frac{1}{q_{m,x} + q_{m,y}^2 - \frac{2i\operatorname{sgn}(q_{m,0})}{\sin\left(\frac{2\pi}{a+1}\right)} |q_{m,0}|^{\frac{2}{a+1}}} \right),
\end{split}
\end{equation}
where $F$ collects all prefactors originating from the individual loop components:
\begin{equation}
    F = c_3^{V/2} \left[ \left(\frac{c_3}{c_2^{a/2}}\right)^{\frac{3}{a+1}} \right]^L \left[ \frac{1}{c_3} \left(\frac{c_3}{c_2^{a/2}}\right)^{\frac{1}{a+1}} \right]^{I_b} \left[ \left(\frac{c_3}{c_2^{a/2}}\right)^{-\frac{2}{a+1}} \right]^{I_f} = c_3^{V/2 - I_b} \left[ \left(\frac{c_3}{c_2^{a/2}}\right)^{\frac{1}{a+1}} \right]^{3L + I_b - 2I_f}.
\end{equation}

The graph topology enforces the exact conservation relations: $2I_f + E_f = 2V$, $2I_b + E_b = V$, and $L = I_b + I_f - V + 1$, where $E_f$ and $E_b$ represent external line counts. Substituting these constraints simplifies the prefactor exactly to:
\begin{equation}
    F = c_3^{E_b/2} \left[ \left(\frac{c_3}{c_2^{a/2}}\right)^{\frac{1}{a+1}} \right]^{-2E_b - \frac{1}{2}E_f + 3}.
\end{equation}
Crucially, $F$ depends strictly on external field configurations ($E_b, E_f$), proving that the flow of $c_2$ and $c_3$ modifies the amplitude uniformly across all perturbation orders without any hidden expansion parameter.


\subsection{Closure of the Fermion Energy Dispersion at Arbitrary Loop Order}

More generally, we extend this formal proof to highly complex, cross-linked (crossed) diagrammatic topologies. For a general $k$-loop fermion self-energy diagram with zero external frequency ($k_0 = 0$), the amplitude contains $k$ loops, $k$ boson lines, and $2k-1$ fermion lines:
\begin{equation}
    kL \sim c_3^k \int \prod_{i=1}^k d^3p_i \left( \prod_{j=1}^{k} \frac{1}{c_2^{a/2}|b_j(p_{i,y}, k_y)|^{a} + c_3 \frac{|c_j(p_{j,0})|}{|d_j(p_{i,y}, k_y)|}} \right) \prod_{l=1}^{2k-1} \frac{1}{e_l(p_{i,x}, k_x) + \left[f_l(p_{i,y}, k_y)\right]^2 - iB\left(\frac{c_3}{c_2^{a/2}}\right)^{\frac{2}{a+1}}|g_l(p_{i,0})|^{\frac{2}{a+1}}},
\end{equation}
where $b_j, c_j, d_j, e_l, f_l$, and $g_l$ are topology-dependent linear combinations. To neutralize the parameters, we perform a homogeneous scaling transformation strictly on the internal frequency components:
\begin{equation}
    p_{i,0} \longrightarrow \left(\frac{c_2^{a/2}}{c_3}\right) p_{i,0}.
\end{equation}
Under this rescaling, the integration measure yields a Jacobian factor of $\left(c_2^{a/2}/c_3\right)^k$. Since every Landau-damping term pulls out a factor of $c_2^{a/2}/c_3$, the amplitude transforms into:
\begin{equation}
\begin{split}
    kL & \sim c_3^k \left(\frac{c_2^{a/2}}{c_3}\right)^k \left(\frac{1}{c_2^{a/2}}\right)^k \int \prod_{i=1}^k d^3p_i \left( \prod_{j=1}^{k} \frac{1}{|b_j(p_{i,y}, k_y)|^{a} + \frac{|c_j(p_{j,0})|}{|d_j(p_{i,y}, k_y)|}} \right) \prod_{l=1}^{2k-1} \frac{1}{e_l(p_{i,x}, k_x) + \left[f_l(p_{i,y}, k_y)\right]^2 - iB|g_l(p_{i,0})|^{\frac{2}{a+1}}} \\
    & = \int \prod_{i=1}^k d^3p_i \left( \prod_{j=1}^{k} \frac{1}{|b_j(p_{i,y}, k_y)|^{a} + \frac{|c_j(p_{j,0})|}{|d_j(p_{i,y}, k_y)|}} \right) \prod_{l=1}^{2k-1} \frac{1}{e_l(p_{i,x}, k_x) + \left[f_l(p_{i,y}, k_y)\right]^2 - iB|g_l(p_{i,0})|^{\frac{2}{a+1}}}.
\end{split}
\end{equation}
All explicit occurrences of $c_2$ and $c_3$ cancel out identically ($c_3^k c_3^{-k} = 1$, $c_2^{ak/2} c_2^{-ak/2} = 1$), proving parameter independence across all topologies. The spatial fields are defined via identical loop coefficients $A_{li}$ as:
\begin{equation}
    e_l(p_{i,x}, k_x) = \chi_l k_x + \sum_i A_{li} p_{i,x}, \qquad f_l(p_{i,y}, k_y) = \chi_l k_y + \sum_i A_{li} p_{i,y}.
\end{equation}

We classify the internal fermion lines using an index $\chi_l \in \{0, 1\}$ to systematically track their dependence on the external momentum:
\begin{itemize}
    \item \textbf{The $\chi_l = 1$ Case (External Momentum-Carrying Lines):} By shifting the internal momenta via $p_{i,x} = \tilde{p}_{i,x} - 2k_y p_{i,y}$, the kinetic term becomes:
    \begin{equation}
        \label{eq:chi_1_momentum} 
        e_l(p_{i,x}, k_x) + \left[ f_l(p_{i,y}, k_y) \right]^2 = \left( k_x + k_y^2 \right) + \sum_i A_{li} \tilde{p}_{i,x} + \left( \sum_i A_{li} p_{i,y} \right)^2.
    \end{equation}
    This explicitly locks the dependence on the external momenta $k_x$ and $k_y$ into the invariant combination $k_x + k_y^2$.
    
    \item \textbf{The $\chi_l = 0$ Case (Pure Internal Loops):} These lines form closed internal fermion loops. The loop momenta flowing through these propagators are entirely independent of the momenta constituting the $\chi_l=1$ lines. And, they do not carry explicit dependence on the external combination $k_x + k_y^2$.
\end{itemize}
In both cases, momentum conservation at the vertices ($p_{\text{out}} - p_{\text{in}} = q$) guarantees that the internal boson lines remain rigorously decoupled from the external fermion momentum $k$ in the fermion self-energy diagrams.

Expanding the amplitude as a power series gives $kL \equiv \mathcal{A} + \mathcal{B}(k_x + k_y^2) + \mathcal{C}(k_x + k_y^2)^2 + \dots$. Since integration occurs within an infinitesimally thin shell near the UV cutoff ($k_x + k_y^2 \ll \text{internal loops}$), we introduce a non-linear scaling transformation onto a primary internal frequency scale $p_{1,\omega}$:
\begin{equation}
    p_{i,x} \longrightarrow \beta_i \left( p_{1,\omega} \right)^{\frac{2}{a+1}}, \qquad p_{i,y} \longrightarrow \gamma_i \left( p_{1,\omega} \right)^{\frac{1}{a+1}}, \qquad p_{i,\omega} \longrightarrow \alpha_i p_{1,\omega} \quad (i \neq 1).
\end{equation}
Factoring out the volume scaling transforms the amplitude series into:
\begin{equation}
    kL \sim \int d\left(p_{1,\omega}\right) p_{1,\omega}^{-1 + \frac{2}{a+1}} \int d\beta_1 d\gamma_1 \int \left( \prod_{i=2}^{k} d\alpha_i d\beta_i d\gamma_i \right) \sum_{n=0}^\infty p_{1,\omega}^{-\frac{2n}{a+1}} (k_x + k_y^2)^n F_n\left( \{\alpha_i, \beta_i, \gamma_i\} \right).
\end{equation}
The physical relevance of each term is determined by its scaling dimensionality:
\begin{enumerate}
    \item \textbf{The $n=0$ Term:} The scale power is $p_{1,\omega}^{-1 + \frac{2}{a+1}}$, yielding a non-universal power-law divergence $ \Lambda_{1,\omega}^{\frac{2}{a+1}}$ that only shifts the unphysical background constant.
    \item \textbf{The $n=1$ Term:} The net exponent is exactly $-1$ (since $-1 + \frac{2}{a+1} - \frac{2}{a+1} = -1$). This isolates a pure logarithmic divergence $\int \frac{d p_{1,\omega}}{p_{1,\omega}} \sim \ln (\Lambda_y / \Lambda_{y,0})$ proportional to $k_x + k_y^2$, which uniquely governs the RG scaling flow of the self-energy.
    \item \textbf{The $n > 1$ Terms:} Integrating higher-order terms generates only negative powers of cutoff.
\end{enumerate}

\subsection{Closure of the Fermion Self-Energy at Arbitrary Loop Order}

We now evaluate the explicit scaling structure of the $k$-loop fermion self-energy amplitude under zero external spatial momentum ($k_x = k_y = 0$) but with a non-vanishing external frequency $k_0$. The diagrammatic skeleton is formally given by:
\begin{equation}
    kL \sim c_3^k \int \prod_{i=1}^k d^3p_i \prod_{j=1}^{k} \frac{1}{c_2^{a/2}|b_j(p_{i,y})|^{a} + c_3 \frac{|c_j(p_{j,0}, k_0)|}{|d_j(p_{j,y})|}} \prod_{l=1}^{2k-1} \frac{1}{e_l(p_{i,x}) + \left[f_l(p_{i,y})\right]^2 - iB\left(\frac{c_3}{c_2^{a/2}}\right)^{\frac{2}{a+1}}|g_l(p_{i,0}, k_0)|^{\frac{2}{a+1}}}.
\end{equation}

To systematically extract the tracking coefficients, we perform the spatial variable transformation $(p_{i,x}, p_{i,y}) \rightarrow \left( \left(\frac{c_3}{c_2^{a/2}}\right)^{\frac{2}{a+1}} p_{i,x}, \left(\frac{c_3}{c_2^{a/2}}\right)^{\frac{1}{a+1}} p_{i,y} \right)$ across all loop indices. Collecting the corresponding Jacobian volume elements and propagator prefactors yields:
\begin{equation}
\begin{split}
    kL & \sim c_3^k \left[ \left(\frac{c_3}{c_2^{a/2}}\right)^{\frac{3}{a+1}} \right]^k \left[ \frac{1}{c_3} \left(\frac{c_3}{c_2^{a/2}}\right)^{\frac{1}{a+1}} \right]^k \left[ \left(\frac{c_3}{c_2^{a/2}}\right)^{-\frac{2}{a+1}} \right]^{2k-1} \\
    & \quad \times \int \prod_{i=1}^k d^3p_i \prod_{j=1}^{k} \frac{1}{|b_j(p_{i,y})|^{a} + \frac{|c_j(p_{j,0}, k_0)|}{|d_j(p_{j,y})|}} \prod_{l=1}^{2k-1} \frac{1}{e_l(p_{i,x}) + \left[f_l(p_{i,y})\right]^2 - iB|g_l(p_{i,0}, k_0)|^{\frac{2}{a+1}}} \\
    & = \left(\frac{c_3}{c_2^{a/2}}\right)^{\frac{2}{a+1}} \int \prod_{i=1}^k d^3p_i \prod_{j=1}^{k} \frac{1}{|b_j(p_{i,y})|^{a} + \frac{|c_j(p_{j,0}, k_0)|}{|d_j(p_{j,y})|}} \prod_{l=1}^{2k-1} \frac{1}{e_l(p_{i,x}) + \left[f_l(p_{i,y})\right]^2 - iB|g_l(p_{i,0}, k_0)|^{\frac{2}{a+1}}}.
\end{split}
\end{equation}

Next, we introduce a non-perturbative scaling mapping that factors out the external frequency scale $k_0$, defined by the coordinate transformations $(p_{i,0}, p_{i,x}, p_{i,y}) \rightarrow \left( |k_0|\alpha_i, |k_0|^{\frac{2}{a+1}}\beta_i, |k_0|^{\frac{1}{a+1}}\gamma_i \right)$. This substitution directly extracts the analytical scaling behavior of the fermion self-energy:
\begin{equation}
    kL \sim \left(\frac{c_3}{c_2^{a/2}}\right)^{\frac{2}{a+1}} |k_0|^{\frac{2}{a+1}} \int \prod_{i=1}^k d\alpha_i d\beta_i d\gamma_i \prod_{j=1}^{k} \frac{1}{|b_j(\gamma_{i})|^{a} + \frac{|c_j(\alpha_i)|}{|d_j(\gamma_i)|}} \prod_{l=1}^{2k-1} \frac{1}{e_l(\beta_{i}) + \left[f_l(\gamma_{i})\right]^2 - iB|g_l(\alpha_i)|^{\frac{2}{a+1}}}.
\end{equation}
This establishes that the full multi-loop fermion self-energy scales rigorously as $|k_0|^{\frac{2}{a+1}}$, fully consistent with our original low-energy RPA action ansatz.


\subsection{Closure of the $|q_y|^a$ term at Arbitrary Loop Order}

We next examine the scaling behavior of the $k$-loop boson self-energy (polarization) diagram under zero external frequency ($q_0 = 0$) but with a finite external momentum $q$. The corresponding skeletal amplitude can be written as:
\begin{equation}
    kL \sim c_3^k \int \prod_{i=1}^k d^3p_i \prod_{j=1}^{k-1} \frac{1}{c_2^{a/2}|b_j(p_{i,y}, q_y)|^a + c_3 \frac{|c_j(p_{i,0})|}{|d_j(p_{i,y}, q_y)|}} \prod_{l=1}^{2k} \frac{1}{e_l(p_{i,x}, q_x) + \left[f_l(p_{i,y}, q_y)\right]^2 - iB\left(\frac{c_3}{c_2^{a/2}}\right)^{\frac{2}{a+1}}|g_l(p_{i,0})|^{\frac{2}{a+1}}}.
\end{equation}

By implementing a homogeneous frequency coordinate transformation $p_{i,0} \rightarrow \left(\frac{c_2^{a/2}}{c_3}\right) p_{i,0}$, the amplitude extracts the bosonic parameter dependencies through:
\begin{equation}
\begin{split}
    kL & \sim c_3^k \left(\frac{c_2^{a/2}}{c_3}\right)^k \int \prod_{i=1}^k d^3p_i \left(\frac{1}{c_2^{a/2}}\right)^{k-1} \prod_{j=1}^{k-1} \frac{1}{|b_j(p_{i,y}, q_y)|^a + \frac{|c_j(p_{i,0})|}{|d_j(p_{i,y}, q_y)|}} \prod_{l=1}^{2k} \frac{1}{e_l(p_{i,x}, q_x) + \left[f_l(p_{i,y}, q_y)\right]^2 - iB|g_l(p_{i,0})|^{\frac{2}{a+1}}} \\
    & = c_2^{a/2} \int \prod_{i=1}^k d^3p_i \prod_{j=1}^{k-1} \frac{1}{|b_j(p_{i,y}, q_y)|^a + \frac{|c_j(p_{i,0})|}{|d_j(p_{i,y}, q_y)|}} \prod_{l=1}^{2k} \frac{1}{e_l(p_{i,x}, q_x) + \left[f_l(p_{i,y}, q_y)\right]^2 - iB|g_l(p_{i,0})|^{\frac{2}{a+1}}}.
\end{split}
\end{equation}

To project out the full external momentum fields, we apply the anisotropic variable transformations $(p_{i,0}, p_{i,x}, p_{i,y}) \rightarrow \left( |q_y|^{a+1}\alpha_i, |q_y|^2\beta_i, |q_y|\gamma_i \right)$. This maps the multi-loop momentum integration volume into a dimensionless form:
\begin{equation}
\begin{split}
    kL & \sim c_2^{a/2} |q_y|^{k(a+4)} |q_y|^{a(-k+1)} |q_y|^{-4k} \\
    & \quad \times \int \prod_{i=1}^k d\alpha_i d\beta_i d\gamma_i \prod_{j=1}^{k-1} \frac{1}{|b_j(\gamma_i, 1)|^a + \frac{|c_j(\alpha_i)|}{|d_j(\gamma_i, 1)|}} \prod_{l=1}^{2k} \frac{1}{e_l\left(\beta_i, \frac{q_x}{|q_y|^2}\right) + \left[f_l(\gamma_i, 1)\right]^2 - iB|g_l(\alpha_i)|^{\frac{2}{a+1}}} \\
    & = c_2^{a/2}|q_y|^a \int \prod_{i=1}^k d\alpha_i d\beta_i d\gamma_i \prod_{j=1}^{k-1} \frac{1}{|b_j(\gamma_i, 1)|^a + \frac{|c_j(\alpha_i)|}{|d_j(\gamma_i, 1)|}} \prod_{l=1}^{2k} \frac{1}{e_l\left(\beta_i, \frac{q_x}{|q_y|^2}\right) + \left[f_l(\gamma_i, 1)\right]^2 - iB|g_l(\alpha_i)|^{\frac{2}{a+1}}} \\
    & \equiv c_2^{a/2}|q_y|^a F\left(\frac{q_x}{|q_y|^2}\right).
\end{split}
\end{equation}
This confirms that the boson polarization is pinned to the overarching dimension $c_2^{a/2}|q_y|^a$, modulated by a scaling function $F$ that depends strictly on the anisotropic ratio $q_x / |q_y|^2$.


\subsection{Reparameterization Invariance Constraints on the Boson Scaling Function}

To fix the functional form of the scaling profile $F\left(q_x / |q_y|^2\right)$, we apply the local reparameterization invariance of the Fermi surface. Let $(\epsilon_x, \epsilon_y)$ denote an arbitrary shift vector along the local patch of the Fermi surface, satisfying the bare dispersion constraint:
\begin{equation}
    \epsilon_x + \epsilon_y^2 = 0.
\end{equation}

Under a local coordinate transformation relative to this new point, the shifted fermion momentum components $(k'_x, k'_y)$ are defined as\cite{Sachdev2010prb}:
\begin{equation}
    k'_y = k_y - \epsilon_y, \qquad k'_x = k_x - \epsilon_x + 2\epsilon_y(k_y - \epsilon_y).
\end{equation}
It is straightforward to verify that the quadratic fermion dispersion remains strictly invariant under this transformation mapping:
\begin{equation}
    k'_x + k'^2_y = k_x - \epsilon_x + 2\epsilon_y(k_y - \epsilon_y) + (k_y - \epsilon_y)^2 = k_x + k_y^2 - (\epsilon_x + \epsilon_y^2) = k_x + k_y^2.
\end{equation}

For the Yukawa vertex interactions, the internal boson field momentum $q$ is fixed by local momentum conservation between the outgoing and incoming fermion states ($q = k_{\text{out}} - k_{\text{in}}$). Under this same patch reparameterization, the boson momentum components map onto:
\begin{equation}
    q'_x = q_x + 2\epsilon_y q_y, \qquad q'_y = q_y.
\end{equation}
Evaluating the anisotropic momentum scaling ratio under this transformed coordinate system yields:
\begin{equation}
    \frac{q'_x}{q'^2_y} = \frac{q_x + 2\epsilon_y q_y}{q_y^2} \neq \frac{q_x}{q_y^2}.
\end{equation}
Because the physical scaling ratio shifts explicitly with the choice of local coordinate origin ($\epsilon_y$), a non-trivial function $F\left(q_x / |q_y|^2\right)$ violates the fundamental reparameterization symmetry of the underlying Fermi surface. Therefore, to ensure complete reparameterization invariance across the field theory, the scaling profile must collapse into a topological constant:
\begin{equation}
    F\left(\frac{q_x}{|q_y|^2}\right) = \text{Constant}.
\end{equation}
This mathematically forces the boson propagator corrections to be purely isotropic in patch-momentum space, validating our closed-form RG ansatz.


\subsection{Closure of the Yukawa Vertex at Arbitrary Loop Order}

Finally, we evaluate the general scaling behavior of the $k$-loop Yukawa vertex correction diagram at zero external fermion and boson momentum ($k = q = 0$). In this kinematical limit, the diagrammatic amplitude contains $k$ independent internal loops, $k$ internal boson lines, and $2k$ internal fermion lines. The amplitude can be formally expressed as:
\begin{equation}
    kL \sim c_3^{k + \frac{1}{2}} \int \prod_{i=1}^k d^3p_i \prod_{j=1}^{k} \frac{1}{c_2^{a/2}|b_j(p_{i,y})|^a + c_3 \frac{|c_j(p_{i,0})|}{|d_j(p_{i,y})|}} \prod_{l=1}^{2k} \frac{1}{e_l(p_{i,x}) + \left[f_l(p_{i,y})\right]^2 - iB\left(\frac{c_3}{c_2^{a/2}}\right)^{\frac{2}{a+1}}|g_l(p_{i,0})|^\frac{2}{a+1}},
\end{equation}
where the additional factor of $c_3^{1/2}$ originates from the single unrenormalized external Yukawa vertex coupling ($g = \sqrt{4\pi c_3}$).

To evaluate the parameter dependence, we execute the homogeneous frequency coordinate transformation on all internal profiles: $p_{i,0} \rightarrow \left(\frac{c_2^{a/2}}{c_3}\right) p_{i,0}$. Under this scaling mapping, the multi-loop integration measure generates a frequency Jacobian factor of $\left(c_2^{a/2}/c_3\right)^k$. Simultaneously, each of the $k$ internal boson propagators pulls out a factor of $c_3^{-1}\left(c_3/c_2^{a/2}\right)$, whereas the $2k$ internal fermion lines absorb the scaling factors through the non-Fermi-liquid self-energy structure. Combining these nested scaling factors yields:
\begin{equation}
\begin{split}
    kL & \sim c_3^{k + \frac{1}{2}} \left(\frac{c_2^{a/2}}{c_3}\right)^k \int \prod_{i=1}^k d^3p_i \left( \frac{1}{c_2^{a/2}} \right)^k \prod_{j=1}^{k} \frac{1}{|b_j(p_{i,y})|^a + \frac{|c_j(p_{i,0})|}{|d_j(p_{i,y})|}} \prod_{l=1}^{2k} \frac{1}{e_l(p_{i,x}) + \left[f_l(p_{i,y})\right]^2 - iB|g_l(p_{i,0})|^\frac{2}{a+1}} \\
    & = c_3^{1/2} \int \prod_{i=1}^k d^3p_i \prod_{j=1}^{k} \frac{1}{|b_j(p_{i,y})|^a + \frac{|c_j(p_{i,0})|}{|d_j(p_{i,y})|}} \prod_{l=1}^{2k} \frac{1}{e_l(p_{i,x}) + \left[f_l(p_{i,y})\right]^2 - iB|g_l(p_{i,0})|^\frac{2}{a+1}}.
\end{split}
\end{equation}

Crucially, all tracking parameters $c_2$ and $c_3$ inside the integrand cancel out identically ($c_3^k c_3^{-k} = 1$ and $c_2^{ak/2} c_2^{-ak/2} = 1$). The final prefactor depends exclusively on the bare coupling dimension $c_3^{1/2} \sim g$. This exact cancellation provides the mathematical proof that the multi-loop Yukawa vertex corrections are completely invariant under the interaction-driven coordinate transformations. 

This completes the comprehensive scaling analysis of Appendix A, rigorously establishing that the closed-form RG ansatz is structurally robust and parameter-independent across all diagrammatic sectors and topological orders.

\section{Detailed Evaluation of Three-Loop Diagrammatic Corrections}
\label{app:three_loop_details}

For the two-loop case, all diagrammatic contributions vanish identically because the poles of the fermion propagators with respect to the $x$-component of the fermion momentum all reside within the same complex half-plane. Therefore, to obtain a non-vanishing contribution at the three-loop level, a given Feynman diagram must explicitly incorporate fermion propagators associated with different patches of the Fermi surface\cite{Sachdev2010prb}. Consequently, only a highly restricted class of diagrams yields non-zero loop corrections to the low-energy effective theory.


\subsection{Three-Loop Aslamazov-Larkin-Type Boson Polarization}

\begin{figure}[htbp]
  \centering
  \begin{tabular}{cc}
    \includegraphics[width=0.38\linewidth]{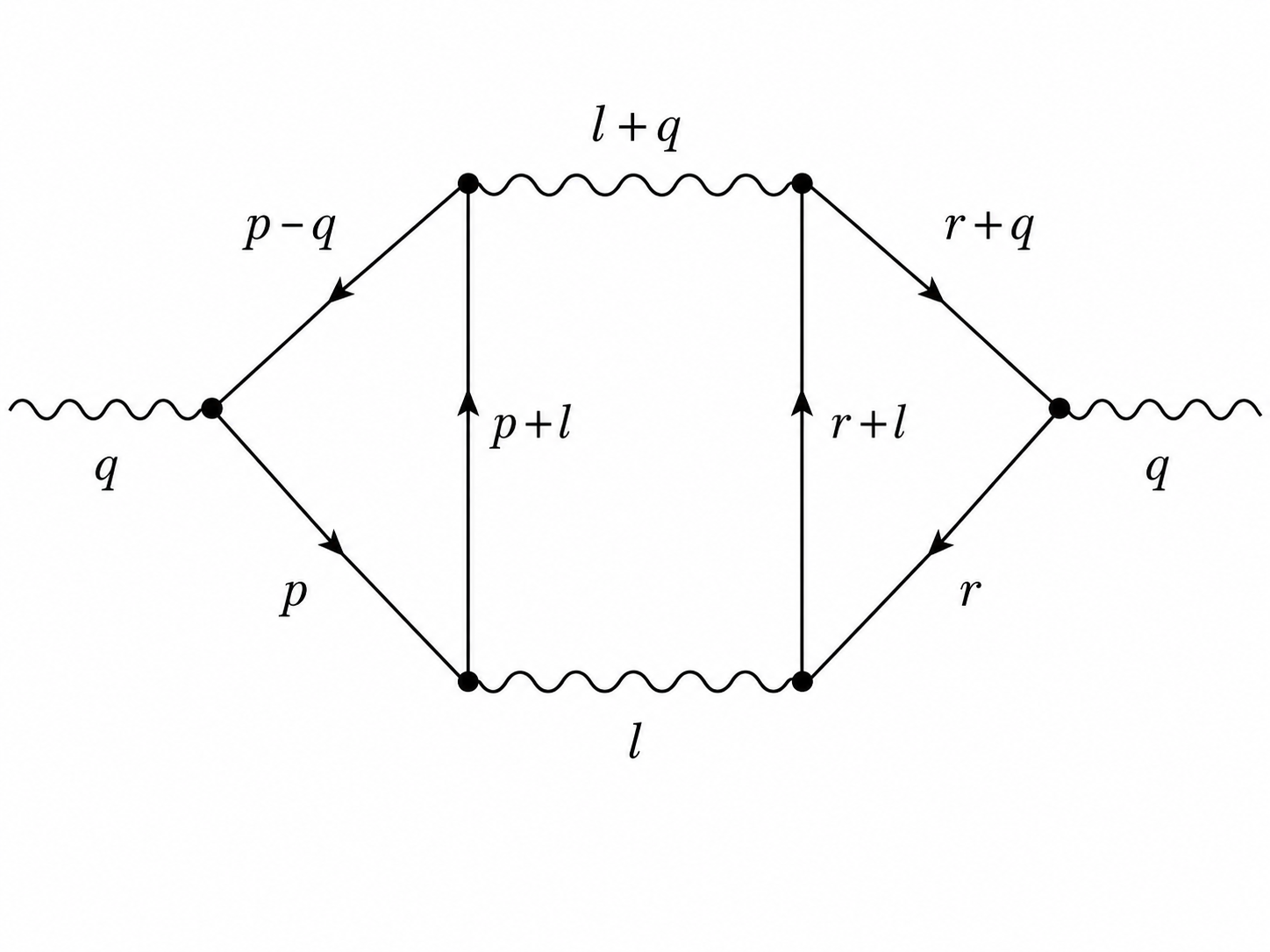} &
    \includegraphics[width=0.4\linewidth]{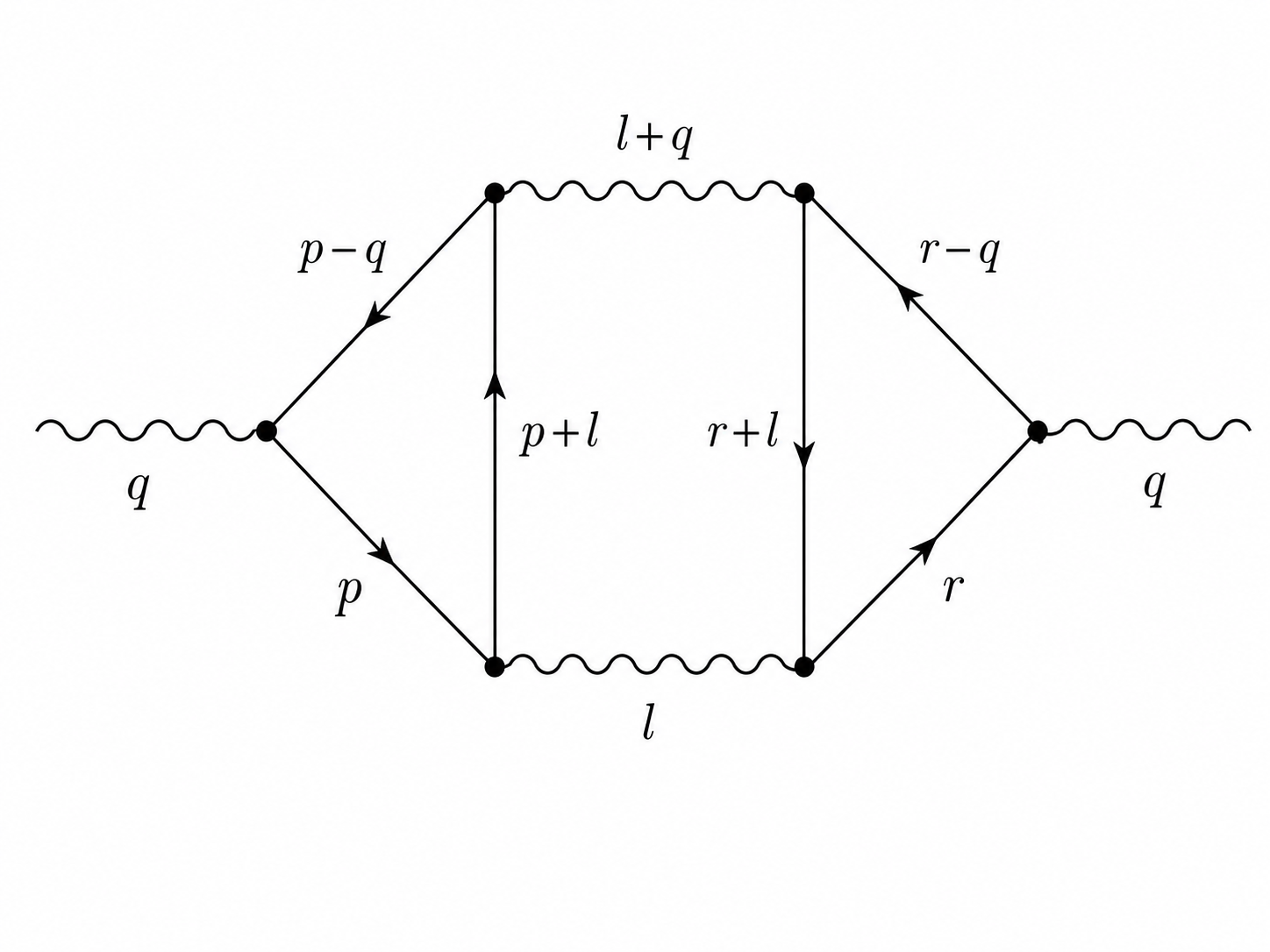} \\
  \end{tabular}
  \caption{Aslamazov-Larkin diagram}
  \label{fig:AL}
\end{figure}

The leading non-vanishing three-loop contribution to the boson self-energy is governed by the Aslamazov-Larkin-type diagrams(Fig. \ref{fig:AL}). At zero external frequency ($q_0 = 0$), the polarization amplitude $\Pi^{3L}(0, \vec{q})$ is formally structured as:
\begin{equation}
\begin{split}
\Pi^{3L}(0, \vec{q}) = &-64\pi^3 c_3^3 \int \frac{d^{3}p d^{3}r d^{3}l}{(2\pi)^9} \frac{1}{c_2^{a/2}|l_y|^a + c_3\frac{|l_0|}{|l_y|}} \frac{1}{c_2^{a/2}|l_y + q_y|^a + c_3\frac{|l_0|}{|l_y+q_y|}} \\
&\times \frac{1}{p_x + p_y^2 - i\Sigma_1(p_0)} \frac{1}{(p_x - q_x) + (p_y - q_y)^2 - i\Sigma_1(p_0)} \frac{1}{(p_x + l_x) + (p_y + l_y)^2 - i\Sigma_1(p_0 + l_0)} \\
&\times \Biggl( \frac{1}{-r_x + r_y^2 - i\Sigma_1(r_0)} \frac{1}{-(r_x - l_x) + (r_y - l_y)^2 - i\Sigma_1(r_0 - l_0)} \frac{1}{-(r_x + q_x) + (r_y + q_y)^2 - i\Sigma_1(r_0)} \\
&\quad + \frac{1}{-r_x + r_y^2 - i\Sigma_1(r_0)} \frac{1}{-(r_x + l_x) + (r_y + l_y)^2 - i\Sigma_1(r_0 + l_0)} \frac{1}{-(r_x - q_x) + (r_y - q_y)^2 - i\Sigma_1(r_0)} \Biggr) \\
&- 64\pi^3 c_3^3 \int \frac{d^{3}p d^{3}r d^{3}l}{(2\pi)^9} \frac{1}{c_2^{a/2}|l_y|^a + c_3\frac{|l_0|}{|l_y|}} \frac{1}{c_2^{a/2}|l_y + q_y|^a + c_3\frac{|l_0|}{|l_y+q_y|}} \\
&\times \frac{1}{-p_x + p_y^2 - i\Sigma_1(p_0)} \frac{1}{-(p_x - q_x) + (p_y - q_y)^2 - i\Sigma_1(p_0)} \frac{1}{-(p_x + l_x) + (p_y + l_y)^2 - i\Sigma_1(p_0 + l_0)} \\
&\times \Biggl( \frac{1}{r_x + r_y^2 - i\Sigma_1(r_0)} \frac{1}{(r_x - l_x) + (r_y - l_y)^2 - i\Sigma_1(r_0 - l_0)} \frac{1}{(r_x + q_x) + (r_y + q_y)^2 - i\Sigma_1(r_0)} \\
&\quad + \frac{1}{r_x + r_y^2 - i\Sigma_1(r_0)} \frac{1}{(r_x + l_x) + (r_y + l_y)^2 - i\Sigma_1(r_0 + l_0)} \frac{1}{(r_x - q_x) + (r_y - q_y)^2 - i\Sigma_1(r_0)} \Biggr),
\end{split}
\end{equation}
where, $\Sigma_1(p)=2\text{sgn}(p_0)\Big(\frac{c_3}{c_2^{a/2}}\Big)^{2/(a+1)}|p_0|^{2/(a+1)}$.
By invoking the loop momentum crossing transformation ($\vec{l} \rightarrow -\vec{l}$ and $p \leftrightarrow r$), the baseline integrations rearrange into a highly condensed symmetry relation:
\begin{equation}
\begin{split}
\Pi^{3L}(0, \vec{q}) = &-64\pi^3 c_3^3 \int \frac{d^{3}p d^{3}r d^{3}l}{(2\pi)^9} \frac{1}{c_2^{a/2}|l_y|^a + c_3\frac{|l_0|}{|l_y|}} \frac{1}{c_2^{a/2}|l_y + q_y|^a + c_3\frac{|l_0|}{|l_y+q_y|}} \\
&\times \frac{1}{p_x + p_y^2 - i\Sigma_1(p_0)} \frac{1}{(p_x - q_x) + (p_y - q_y)^2 - i\Sigma_1(p_0)} \frac{1}{(p_x + l_x) + (p_y + l_y)^2 - i\Sigma_1(p_0 + l_0)} \\
&\times \Biggl( \frac{1}{-r_x + r_y^2 + i\Sigma_1(r_0)} \frac{1}{-(r_x - l_x) + (r_y - l_y)^2 + i\Sigma_1(r_0 + l_0)} \frac{1}{-(r_x + q_x) + (r_y + q_y)^2 + i\Sigma_1(r_0)} \\
&\quad + \frac{1}{-r_x + r_y^2 - i\Sigma_1(r_0)} \frac{1}{-(r_x + l_x) + (r_y + l_y)^2 - i\Sigma_1(r_0 + l_0)} \frac{1}{-(r_x - q_x) + (r_y - q_y)^2 - i\Sigma_1(r_0)} \Biggr) \\
&+ (q \rightarrow -q).
\end{split}
\end{equation}

Executing the contour integrations sequentially over the spatial momentum components $p_x$ and $r_x$ reveals a strict kinematic phase-space boundary governed by step functions $\Theta(-xy)$. Mapping the frequency variables linearly for the $l_0 > 0$ and $l_0 < 0$ domains, the total real part of the amplitude combines with its complex conjugate (c.c.) to yield:
\begin{equation}
    \begin{split}
        \Pi^{3L}(0,\vec q)&=\frac{4c_3^3}{\pi q_y^2}\int _0^{\infty}dl_0\int ^{|q_y|}_0dl_y\int^{l_0}_0dr_0\int ^{l_0}_0dp_0\ \frac{1}{c_2^{a/2}l_y^a+c_3\frac{l_0}{l_y}}\frac{1}{c_2^{a/2}(l_y-|q_y|)^a+c_3\frac{l_0}{l_y-|q_y|}}\\
        &\times\bigg[\frac{1}{C[(l_0-p_0)^{2/(a+1)}+(l_0-r_0)^{2/(a+1)}+(p_0)^{2/(a+1)}+(r_0)^{2/(a+1)}]}\\
        &-\frac{C[(l_0-p_0)^{2/(a+1)}+(l_0-r_0)^{2/(a+1)}+(p_0)^{2/(a+1)}+(r_0)^{2/(a+1)}]}{C^2[(l_0-p_0)^{2/(a+1)}+(l_0-r_0)^{2/(a+1)}+(p_0)^{2/(a+1)}+(r_0)^{2/(a+1)}]^2+4l_y^2(l_y-|q_y|)^2}\bigg],
    \end{split}
\end{equation}
where, $C=\frac{2}{\sin(\frac{2\pi}{a+1})}\Big(\frac{c_3}{c_2^{a/2}}\Big)^{2/(a+1)}$. To systematically factor out the parameter scaling, we implement the dimensionless variable substitutions: $(l_0,p_0,r_0,l_y)\rightarrow (|q_y|^{a+1}\alpha,|q_y|^{a+1}\beta,|q_y|^{a+1}\gamma,|q_y|\epsilon)$. Then, we obtain the scale-invariant form:
\begin{equation}
\Pi^{3L}(0, \vec{q}) = c_2^{a/2} |q_y|^a \times I(a),
\end{equation}
where $I(a)$ is a parameter-independent universal integration profile defined as:
\begin{equation}
\begin{split}
I(a) = \frac{4}{\pi} \int_{0}^{\infty} d\alpha \int_{0}^{\alpha} d\beta \int_{0}^{\alpha} d\gamma \int_{0}^{1} d\epsilon \, &\left( \frac{1}{\epsilon^a + \frac{\alpha}{\epsilon}} \right) \left( \frac{1}{(1 - \epsilon)^a + \frac{\alpha}{1 - \epsilon}} \right) \\
\times \Biggl[ &\frac{\sin\left(\frac{2\pi}{a+1}\right)}{2\left[(\alpha-\beta)^{\frac{2}{a+1}} + (\alpha-\gamma)^{\frac{2}{a+1}} + \beta^{\frac{2}{a+1}} + \gamma^{\frac{2}{a+1}}\right]} \\
&- \frac{2\sin\left(\frac{2\pi}{a+1}\right)\left[(\alpha-\beta)^{\frac{2}{a+1}} + (\alpha-\gamma)^{\frac{2}{a+1}} + \beta^{\frac{2}{a+1}} + \gamma^{\frac{2}{a+1}}\right]}{4\left[(\alpha-\beta)^{\frac{2}{a+1}} + (\alpha-\gamma)^{\frac{2}{a+1}} + \beta^{\frac{2}{a+1}} + \gamma^{\frac{2}{a+1}}\right]^2 + 4\sin^2\left(\frac{2\pi}{a+1}\right)\epsilon^2(\epsilon-1)^2} \Biggr].
\end{split}
\end{equation}
\begin{figure}[htbp]

    \includegraphics[width=0.6\linewidth]{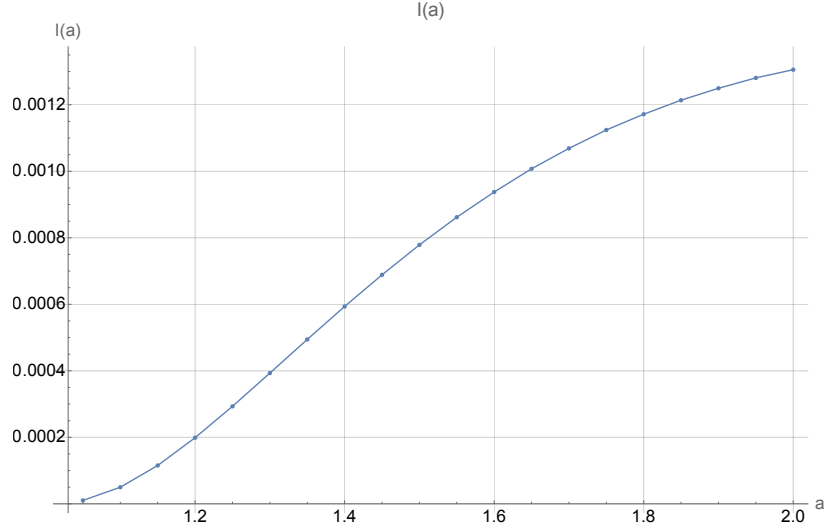}
    \caption{Monte Carlo integration of $I(a)$}
  \label{fig:AL_cal}
\end{figure}

\subsection{Three-Loop Fermion Self-Energy Corrections}
\begin{figure}[htbp]
  \centering
  \begin{tabular}{cc}
    \includegraphics[width=0.45\linewidth]{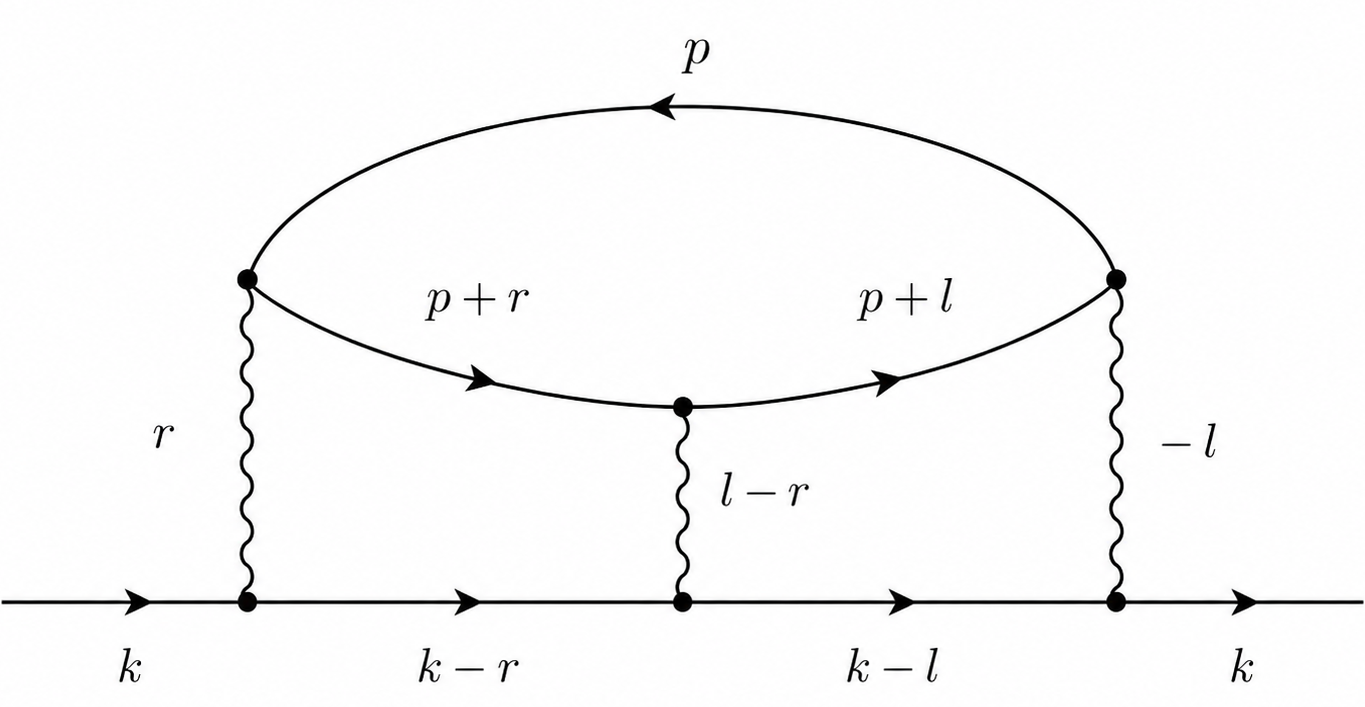} &
    \includegraphics[width=0.43\linewidth]{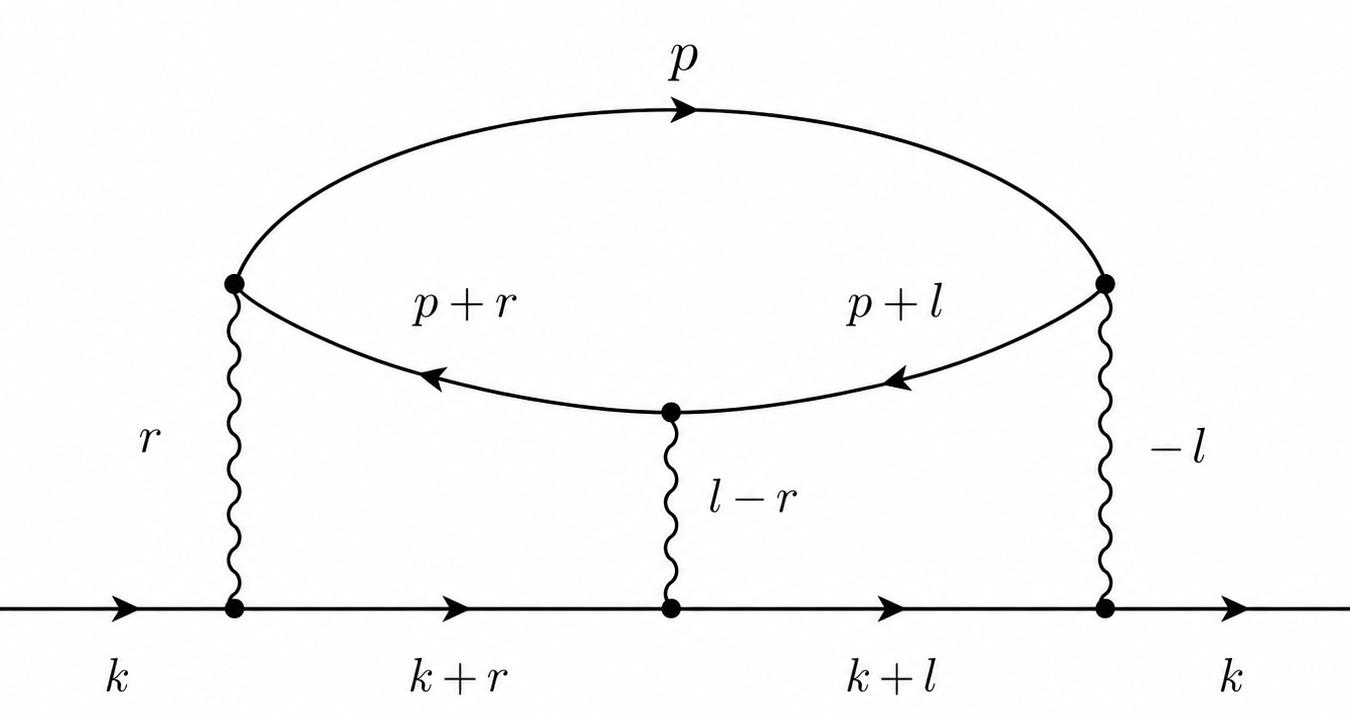} \\
    (a) $\Sigma_1^{3L}$ & (b) $\Sigma_2^{3L}$
  \end{tabular}
 \caption{3-loop fermion self energy that gives log divergence diagrams}
  \label{fig:3-loop_fermion_self_energy}
\end{figure}

Next, we evaluate the distinct multiloop diagrams contributing to the three-loop fermion self-energy(Fig.\ref{fig:3-loop_fermion_self_energy}), parameterized as $\Sigma^{3L}_1(0, \vec{k})$ and $\Sigma^{3L}_2(0, \vec{k})$. The skeletal structure for the primary patch-crossing configuration is written as:
\begin{equation}
    \begin{split}
        \Sigma^{3L}_1(0,\vec k)=&-64\pi^3c_3^3\int\frac{d^3pd^3rd^3l}{(2\pi)^9}\frac{1}{c_2^{a/2}|r_y|^{a}+c_3\frac{|r_0|}{|r_y|}}\frac{1}{c_2^{a/2}|l_y|^{a}+c_3\frac{|l_0|}{|l_y|}}\frac{1}{c_2^{a/2}|r_y-l_y|^{a}+c_3\frac{|r_0-l_0|}{|r_y-l_y|}}\\
        &\ \ \ \ \ \ \ \ \times \frac{1}{(k_x-r_x)+(k_y-r_y)^2-i\Sigma_1(-r)}\frac{1}{(k_x-l_x)+(k_y-l_y)^2-i\Sigma_1(-l)}\\
        & \ \ \ \ \ \ \ \ \times \frac{1}{-p_x+p_y^2-i\Sigma_1(p)}\frac{1}{-(p_x+r_x)+(p_y+r_y)^2-i\Sigma_1(p+r)}\\
        & \ \ \ \ \ \ \ \ \times \frac{1}{-(p_x+l_x)+(p_y+l_y)^2-i\Sigma_1(p+l)}\\
        &=\frac{c_3^3}{2\pi^4}\int d^3p dr_0dr_ydl_0dl_y\frac{1}{c_2^{a/2}|r_y|^{a}+c_3\frac{|r_0|}{|r_y|}}\frac{1}{c_2^{a/2}|l_y|^{a}+c_3\frac{|l_0|}{|l_y|}}\frac{1}{c_2^{a/2}|r_y-l_y|^{a}+c_3\frac{|r_0-l_0|}{|r_y-l_y|}}\\
        & \ \ \ \ \ \ \ \ \ \times \frac{1}{-p_x+p_y^2-iC\{p\}}\frac{\Theta(r_0+p_0)-\Theta(-r_0)}{-p_x+p_y^2-(k_x+k_y^2)+2r_y(k_y+p_y)-i[\Sigma_1(r+p)+\Sigma_1(r)]}\\
        & \ \ \ \ \ \ \ \ \ \times \frac{\Theta(l_0+p_0)-\Theta(-l_0)}{-p_x+p_y^2-(k_x+k_y^2)+2l_y(k_y+p_y)-i[\Sigma_1(l+p)+\Sigma_1(l)]},
    \end{split}
\end{equation}
where, $\Sigma_1(p)=2\text{sgn}(p_0)\Big(\frac{c_3}{c_2^{a/2}}\Big)^{2/(a+1)}|p_0|^{2/(a+1)}$. By separating the poles across the frequency domains, the non-vanishing components reduce to symmetric combinations under the branch transformation $(l_0, l_y) \leftrightarrow (r_0, r_y)$. Shifting the loop variable $p_y \rightarrow p_y - k_y$ isolates the spatial configurations cleanly:
\begin{equation}
    \begin{split}
        \Sigma^{3L}_1(0,\vec k)&=-\frac{4c_3^3}{\pi^2}\int^\infty_0dp_0\int_{p_0}^{\infty}dr_0\int^{\infty}_{0}dl_0\int^{
        \infty}_0 dr_y\int_{r_y}^\infty dl_y \frac{1}{c_2^{a/2}r_y^{a}+c_3\frac{r_0}{r_y}} \frac{1}{c_2^{a/2}l_y^{a}+c_3\frac{l_0}{l_y}}\frac{1}{c_2^{a/2}(l_y-r_y)^{a}+c_3\frac{r_0+l_0}{l_y-r_y}}\\
        &\times \frac{1}{(l_y-r_y)(k_x+k_y^2)-iC\{r_y[l_0^{2/(a+1)}+(l_0+p_0)^{2/(a+1)}-p_0^{2/(a+1)}]+l_y[p_0^{2/(a+1)}+r_0^{2/(a+1)}+(r_0-p_0)^{2/(a+1)}]\}}\\
        &-\frac{2c_3^3}{\pi^2}\int^{\infty}_0dp_0\int^{\infty}_{p_0}dr_0\int^{\infty}_{p_0}dl_0\int^{\infty}_0dr_y\int^{\infty}_0 dl_y\frac{1}{c_2^{a/2}r_y^{a}+c_3\frac{r_0}{r_y}}\frac{1}{c_2^{a/2}l_y^{a}+c_3\frac{l_0}{l_y}}\frac{1}{c_2^{a/2}(l_y+r_y)^{a}+c_3\frac{|r_0-l_0|}{l_y+r_y}}\\
        &\times \frac{1}{(r_y+l_y)(k_x+k_y^2)-iC\{r_y[(l_0-p_0)^{2/(a+1)}+l_0^{2/(a+1)}+p_0^{2/(a+1)}]+l_y[(r_0-p_0)^{2/(a+1)}+r_0^{2/(a+1)}+p_0^{2/(a+1)}]\}}\\
        &+\text{h.c.}.
    \end{split}
\end{equation}
We factor out the infrared scaling by introducing the scale-invariant mappings: \\
$(p_0,r_0,l_0,r_y,l_y)\rightarrow\bigg(\big(\frac{c_2^{a/2}}{c_3}\big)p_0,\big(\frac{c_2^{a/2}}{c_3}\big)p_0\alpha,\big(\frac{c_2^{a/2}}{c_3}\big)p_0\beta,p_0^{1/(a+1)}\epsilon,p_0^{1/(a+1)}\omega\bigg)$. Taylor expanding the energy denominator in powers of the external invariant momentum $k_x + k_y^2$ directly extracts the dominant logarithmic divergence:
\begin{equation}
\Sigma^{3L}_1(0, \vec{k}) = (k_x + k_y^2) \left( \mathcal{J}_1(a) + \mathcal{J}_2(a) \right) \ln \Lambda_0,
\end{equation}
with the explicit functional profiles mapping as:
\begin{equation}
\begin{split}
\mathcal{J}_1(a)&=-\frac{2}{\pi^2}\sin^2\Big(\frac{2\pi}{a+1}\Big)\int_{1}^{\infty}d\alpha\int^{\infty}_{0}d\beta\int^{
        \infty}_0 d\epsilon\int_{\epsilon}^\infty d\omega \frac{1}{\epsilon^{a}+\frac{\alpha}{\epsilon}} \frac{1}{\omega^{a}+\frac{\beta}{\omega}}\frac{1}{(\omega-\epsilon)^{a}+\frac{\alpha+\beta}{\omega-\epsilon}}\\
        &\times \frac{(\omega-\epsilon)}{\{\epsilon[\beta^{2/(a+1)}+(\beta+1)^{2/(a+1)}-1]+\omega[\alpha^{2/(a+1)}+(\alpha-1)^{2/(a+1)}+1]\}^2},
 \end{split}
\end{equation}
\begin{equation}
\begin{split}
\mathcal{J}_2(a)=&-\frac{1}{\pi^2}\sin^2\Big(\frac{2\pi}{a+1}\Big)\int^{\infty}_{1}d\alpha\int^{\infty}_{1}d\beta\int^{\infty}_0d\epsilon\int^{\infty}_0 d\omega \frac{1}{\epsilon^{a}+\frac{\alpha}{\epsilon}}\frac{1}{\omega^{a}+\frac{\beta}{\omega}}\frac{1}{(\epsilon+\omega)^{a}+\frac{|\alpha-\beta|}{\epsilon+\omega}}\\
        &\times \frac{(\omega+\epsilon)}{\{\epsilon[(\beta-1)^{2/(a+1)}+\beta^{2/(a+1)}+1]+\omega[(\alpha-1)^{2/(a+1)}+\alpha^{2/(a+1)}+1]\}^2},
 \end{split}
\end{equation}
\noindent A parallel calculation for the vertex-dressed self-energy skeleton $\Sigma^{3L}_2(0, \vec{k})$ yields an identical logarithmic tracking flow:
\begin{equation}
\Sigma^{3L}_2(0, \vec{k}) = \Sigma^{3L}_2(0, 0) + (k_x + k_y^2) \left( \mathcal{J}_3(a) + \mathcal{J}_4(a) \right) \ln \Lambda_0,
\end{equation}
with
\begin{equation}
\begin{split}
\mathcal{J}_3(a)&=-\frac{2}{\pi^2}\int^\infty_1 d\alpha\int^\infty_0d\beta\int^\infty_0d\epsilon\int^\infty_\epsilon d\omega \frac{1}{\epsilon^a+\frac{\alpha}{\epsilon}}\frac{1}{\omega^a+\frac{\beta}{\omega}}\frac{1}{(\omega-\epsilon)^a+\frac{\alpha+\beta}{\omega-\epsilon}}\\
\ \ \ \ \ \ \ \ \ \ \ &\times \frac{(\omega-\epsilon)\Big\{\epsilon^2\omega^2(\omega-\epsilon)^2-\frac{1}{\sin^2(\frac{2\pi}{a+1})}[\epsilon[(\beta+1)^{\frac{2}{a+1}}+\beta^{\frac{2}{(a+1)}}-1]+\omega[(\alpha-1)^{\frac{2}{(a+1)}}+\alpha^{\frac{2}{(a+1)}}+1]]^2\Big\}}{\Big\{\epsilon^2\omega^2(\omega-\epsilon)^2+\frac{1}{\sin^2(\frac{2\pi}{a+1})}\{\epsilon[(\beta+1)^{\frac{2}{a+1}}+\beta^{\frac{2}{(a+1)}}-1]+\omega[(\alpha-1)^{\frac{2}{(a+1)}}+\alpha^{\frac{2}{(a+1)}}+1]\}^2\Big\}^2},
\end{split}
\end{equation}
and
\begin{equation}
    \begin{split}
        \mathcal{J}_4(a)&=-\frac{1}{\pi^2}\int^\infty_1 d\alpha\int^\infty_1d\beta\int^\infty_0d\epsilon\int^\infty_0 d\omega \frac{1}{\epsilon^a+\frac{\alpha}{\epsilon}}\frac{1}{\omega^a+\frac{\beta}{\omega}}\frac{1}{(\omega+\epsilon)^a+\frac{|\alpha-\beta|}{\omega+\epsilon}}\\
\ \ \ \ \ \ \ \ \ \ \ &\times \frac{(\omega+\epsilon)\Big\{\epsilon^2\omega^2(\omega+\epsilon)^2-\frac{1}{\sin^2(\frac{2\pi}{a+1})}[\epsilon[(\beta-1)^{\frac{2}{a+1}}+\beta^{\frac{2}{(a+1)}}+1]+\omega[(\alpha-1)^{\frac{2}{(a+1)}}+\alpha^{\frac{2}{(a+1)}}+1]]^2\Big\}}{\Big\{\epsilon^2\omega^2(\omega+\epsilon)^2+\frac{1}{\sin^2(\frac{2\pi}{a+1})}\{\epsilon[(\beta-1)^{\frac{2}{a+1}}+\beta^{\frac{2}{(a+1)}}+1]+\omega[(\alpha-1)^{\frac{2}{(a+1)}}+\alpha^{\frac{2}{(a+1)}}+1]\}^2\Big\}^2}
    \end{split}
\end{equation}
Figure.~\ref{fig:Jtotal} shows the $a$-dependence of
$\mathcal{J}_1(a),\ldots,\mathcal{J}_4(a)$ for $1\le a\le2$.
Summing over all configurations confirms that the total three-loop
fermion self-energy correction scales homogeneously with the external
dispersion:
\begin{equation}
\Sigma^{3L}(0, \vec{k}) = (k_x + k_y^2) \left( \mathcal{J}_1(a) + \mathcal{J}_2(a) + \mathcal{J}_3(a) + \mathcal{J}_4(a) \right) \ln \Lambda_0=(k_x + k_y^2) \mathcal{J}_{total}(a)\ln\Lambda_0.
\end{equation}
\begin{figure}[htbp]
  \centering
  \includegraphics[width=0.72\linewidth]{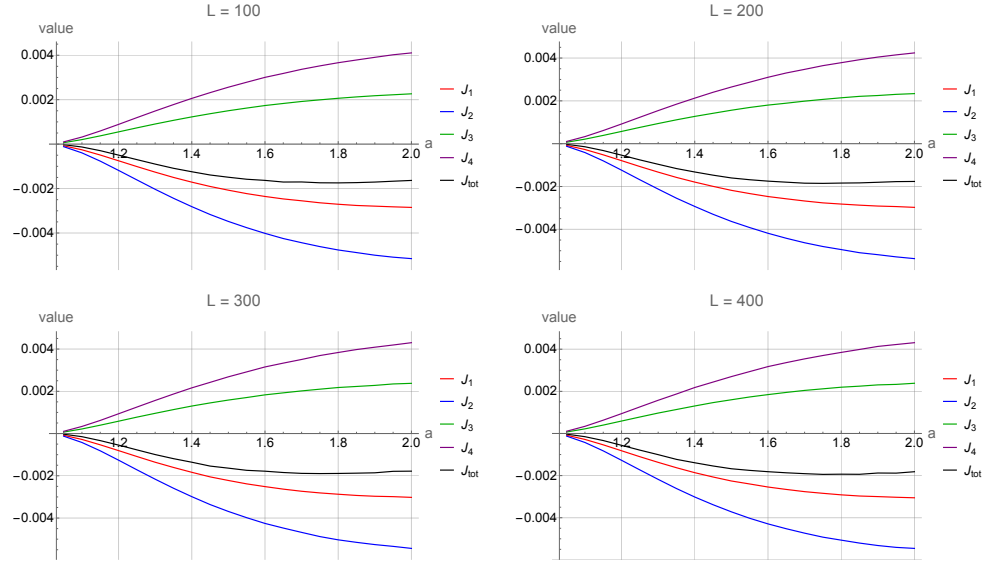}
  \caption{$a$-dependence of the dimensionless coefficients $\mathcal{J}_1(a),\ldots,\mathcal{J}_4(a)$ and their sum $\mathcal{J}_{\rm tot}(a)=\sum_{i=1}^4\mathcal{J}_i(a)$ for $1\le a\le 2$}
  \label{fig:Jtotal}
\end{figure}


\subsection{Three-Loop Yukawa Vertex Corrections}

\begin{figure}[htbp]
  \centering
  \begin{tabular}{cc}
    \includegraphics[width=0.43\linewidth]{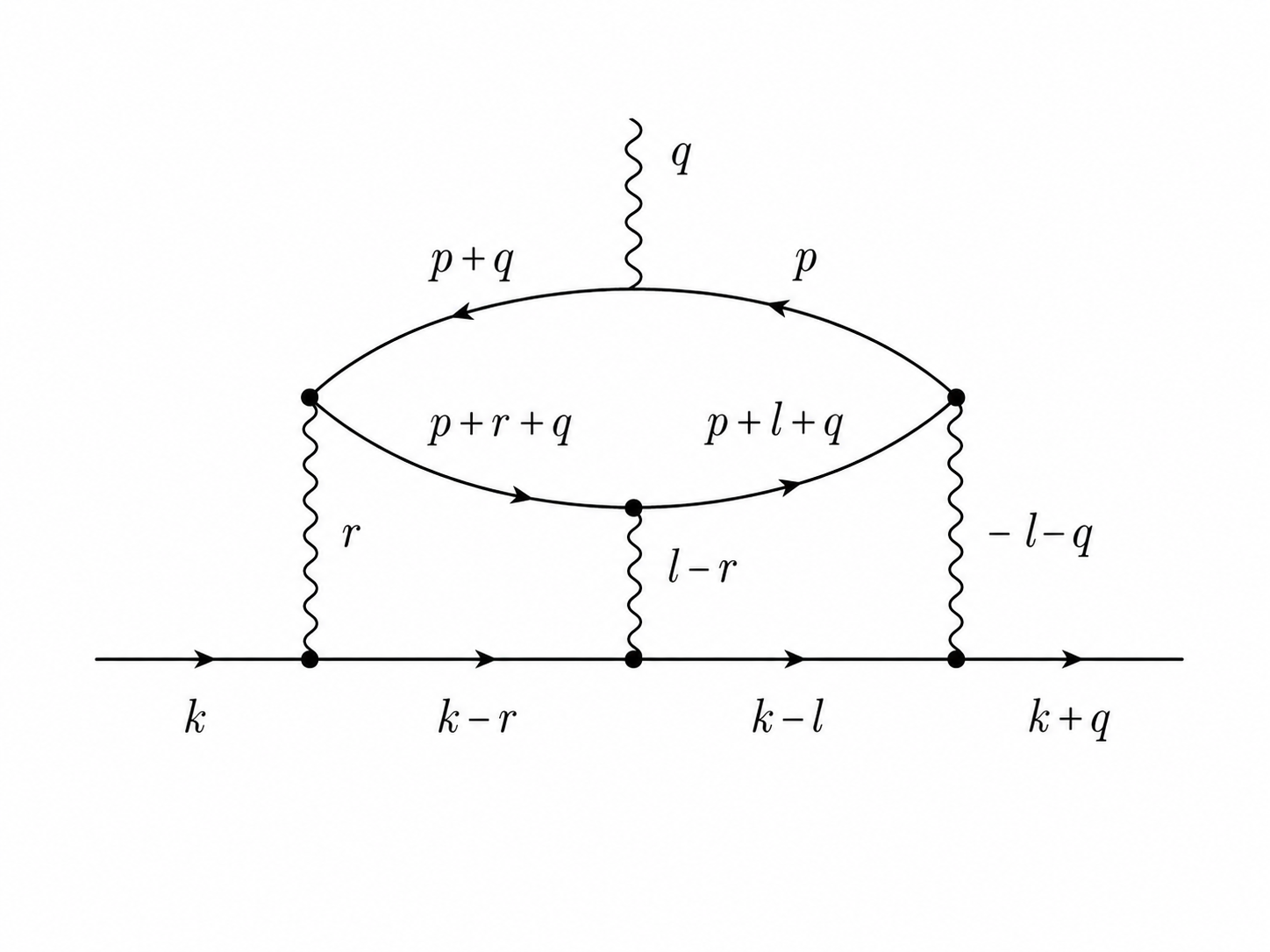} &
    \includegraphics[width=0.45\linewidth]{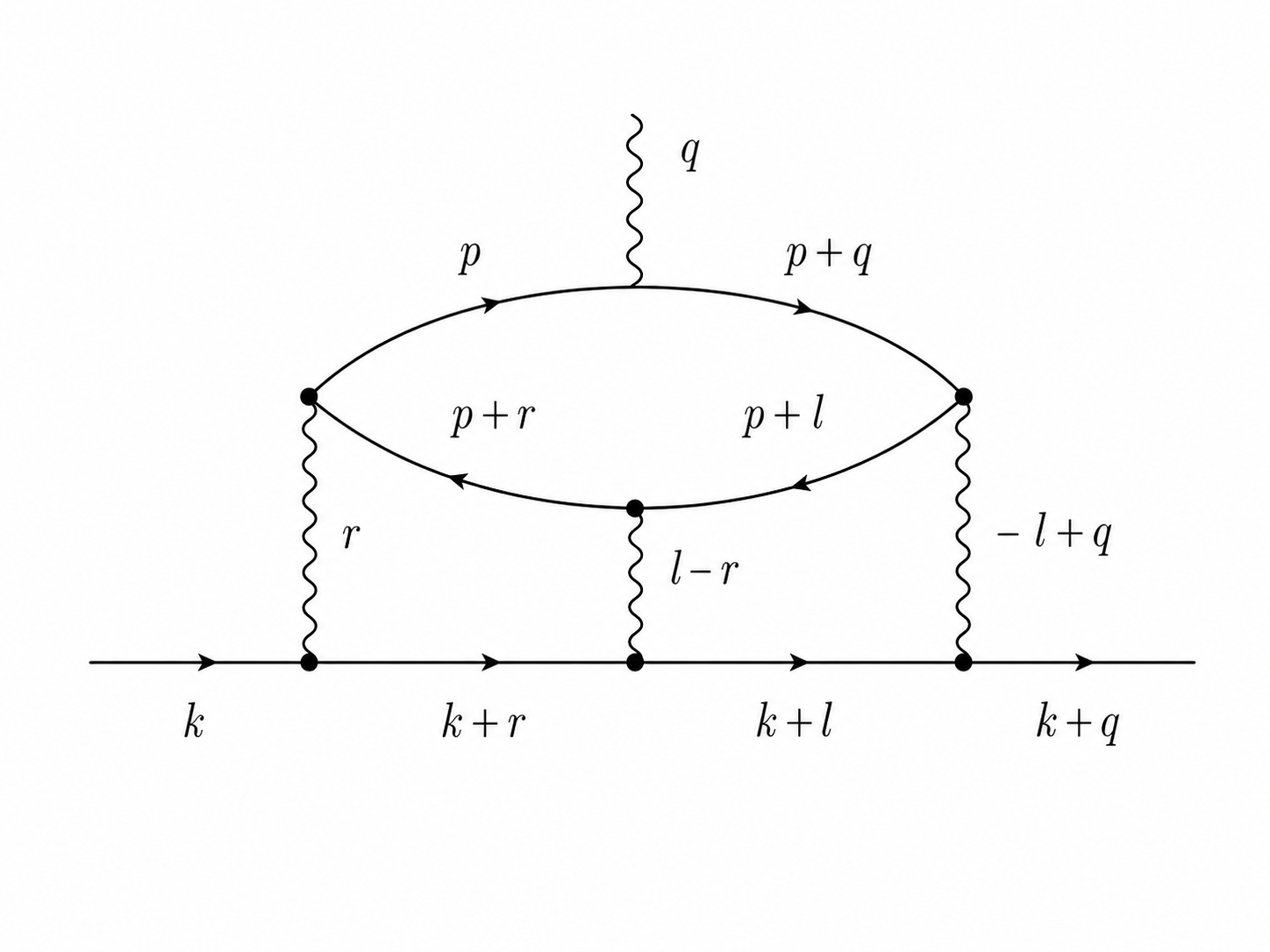} \\
    (a) $\Gamma^{3L}_1$ & (b) $\Gamma^{3L}_2$
  \end{tabular}
  \caption{3-loop vertex diagram that gives log divergence diagrams}
  \label{fig:3-loop_vertex_correction}
\end{figure}

Finally, we evaluate the corresponding three-loop vertex corrections(Fig.\ref{fig:3-loop_vertex_correction}) $\Gamma^{3L}_1(0,0)$ and $\Gamma^{3L}_2(0,0)$ at zero external momentum. Combining the nested matrix element insertions and integrating out the spatial coordinates normal to the Fermi surface transforms the first vertex skeleton into:
\begin{equation}
\begin{split}
    \Gamma^{3L}_1(0,0)&=\sqrt{4\pi c_3}\frac{4c_3^3}{\pi^2}\int^\infty_0dp_0\int_{p_0}^{\infty}dr_0\int^\infty_0dl_0\int^\infty_0 dr_y\int ^{\infty}_{r_y}dl_y \frac{1}{c_2^{a/2}r_y^{a}+c_3\frac{r_0}{r_y}} \frac{1}{c_2^{a/2}l_y^{a}+c_3\frac{l_0}{l_y}}\frac{1}{c_2^{a/2}(l_y-r_y)^{a}+c_3\frac{l_0+r_0}{l_y-r_y}} \\
    & \frac{(l_y-r_y)}{C^2\Big[r_y[(l_0+p_0)^{\frac{2}{a+1}}+l_0^{\frac{2}{a+1}}-p_0^{\frac{2}{a+1}}]+l_y[p_0^{\frac{2}{a+1}}+r_0^{\frac{2}{a+1}}+(r_0-p_0)^{\frac{2}{a+1}}]\Big]^2}\\
        &+\sqrt{4\pi c_3}\frac{2c_3^3}{\pi^2}\int^\infty_0dp_0\int_{p_0}^{\infty}dr_0\int^{\infty}_{p_0}dl_0\int^\infty_0 dr_y\int ^\infty _0 dl_y \frac{1}{c_2^{a/2}|r_y|^{a}+c_3\frac{r_0}{|r_y|}}\frac{1}{c_2^{a/2}|l_y|^{a}+c_3\frac{l_0}{|l_y|}}\frac{1}{c_2^{a/2}|r_y+l_y|^{a}+c_3\frac{|r_0-l_0|}{|r_y+l_y|}}\\
        & \ \ \frac{(l_y+r_y)}{C^2\Big[r_y[(l_0-p_0)^{\frac{2}{a+1}}+l_0^{\frac{2}{a+1}}+p_0^{\frac{2}{a+1}}]+l_y[p_0^{\frac{2}{a+1}}+r_0^{\frac{2}{a+1}}+(r_0-p_0)^{\frac{2}{a+1}}]\Big]^2}\\
        &  +h.c..
\end{split}
\end{equation}
By implementing the identical scale-invariant mapping 
$$(p_0,r_0,l_0,r_y,l_y)\rightarrow\bigg(\big(\frac{c_2^{a/2}}{c_3}\big)p_0,\big(\frac{c_2^{a/2}}{c_3}\big)p_0\alpha,\big(\frac{c_2^{a/2}}{c_3}\big)p_0\beta,p_0^{1/(a+1)}\epsilon,p_0^{1/(a+1)}\omega\bigg),$$ the frequency integral factors out a pure logarithmic infrared divergence:
\begin{equation}
\Gamma^{3L}_1(0, 0) = \sqrt{4\pi c_3} \left( \mathcal{F}_1(a) + \mathcal{F}_2(a) \right) \ln \Lambda_0,
\end{equation}
And, exactly same calculation procedure gives
\begin{equation}
\Gamma^{3L}_2(0, 0) = \sqrt{4\pi c_3} \left( \mathcal{F}_3(a) + \mathcal{F}_4(a) \right) \ln \Lambda_0.
\end{equation}
where,

\begin{equation}
\mathcal{J}_1(a) = -\mathcal{F}_1(a), \qquad \mathcal{J}_2(a) = -\mathcal{F}_2(a), \qquad \mathcal{J}_3(a) = \mathcal{F}_3(a), \qquad \mathcal{J}_4(a) = \mathcal{F}_4(a).
\end{equation}
Summing the total vertex contributions yields the closed-form expression:
\begin{equation}
\Gamma^{3L}(0, 0) = \sqrt{4\pi c_3} \left( -\mathcal{J}_1(a) - \mathcal{J}_2(a) + \mathcal{J}_3(a) + \mathcal{J}_4(a) \right) \ln \Lambda_0.
\end{equation}
This exact algebraic matching completes the full derivation of the three-loop sector, establishing the technical baseline for the fixed-point consistency conditions discussed in the main text.
\end{widetext}


\end{document}